\documentclass{article}

\usepackage{arxiv}

\usepackage{xcolor}
\definecolor{kitfarbe}{rgb}{0.004,0.588,0.51}
\definecolor{urlcol}{HTML}{294E85}
\definecolor{anccol}{HTML}{B52D26}  

\usepackage[utf8]{inputenc}
\usepackage[T1]{fontenc}

\usepackage{booktabs}  
\usepackage{longtable}
\usepackage{amsfonts}  
\usepackage{nicefrac}  

\usepackage{microtype}  
\usepackage{graphicx}
\usepackage{caption}
\usepackage{tabularx}
\usepackage{makecell}

\usepackage{amsmath}
\usepackage{mathtools}
\usepackage{siunitx}
\usepackage[sort&compress]{natbib}
\graphicspath{ {./figures/} }
\usepackage[colorlinks=true,
            urlcolor=urlcol,
            linkcolor=anccol,
            citecolor=kitfarbe]{hyperref}

\title{Automated Detection of Match Phases in Football from Spatio-Temporal Tracking Data Using Graph Neural Networks}

\author{
  Vincent Renner\textsuperscript{1,2} \quad Nils Koster\textsuperscript{1,3} \quad
  Pascal Bauer\textsuperscript{4,5} \quad Melanie Schienle\textsuperscript{1} \\[6pt]
  \textsuperscript{1}Institute of Statistics, Karlsruhe Institute of Technology, Karlsruhe, Germany\\
  \textsuperscript{2}Karlsruher Sport-Club, Karlsruhe, Germany\\
  \textsuperscript{3}Broad Institute of MIT \& Harvard, Cambridge, MA, USA\\
  \textsuperscript{4}German Football Association (DFB), Frankfurt, Germany\\
  \textsuperscript{5}Chair for Sports Analytics, Saarland University, Saarbrücken, Germany\\[4pt]
\texttt{\{vincent.renner,nils.koster,melanie.schienle\}@kit.edu}\\
\texttt{pascal.bauer@dfb.de}
}

\begin{document}
\maketitle
\begin{abstract}
	Spatio-temporal tracking data has opened new possibilities for detecting complex tactical patterns in football, yet modeling the interactive movements of multiple players remains challenging.
	This paper proposes a framework combining graph neural networks (GNNs) with a sequential model to classify match phases on a second-by-second basis across a seven-class taxonomy. Match phase classification is tactically meaningful, and the availability of rule-based labels across 203 matches makes it a suitable testbed for a systematic comparison of adjacency constructions and message-passing layers, a question that has received limited attention in existing research. 
	Our selected GNN-LSTM model outperforms all aggregated-feature baselines, including XGBoost and a Long Short-Term Memory (LSTM) network, as the strongest baseline scores 4.6\% lower in macro F1. Graph representations using a domain-informed Delaunay triangulation that approximates passing lanes, paired with a custom Spatial Edge-Augmented Convolution (SEAConv) layer that injects edge attributes directly into messages, achieve the best performance by capturing spatial dependencies while limiting uninformative messages from redundant edges. 
	Integrated Gradients attributions indicate the model uses spatial player configurations, particularly horizontal positioning, in combination with possession and ball-status indicators. 
	This work offers an automated solution for fine-grained tactical analysis, reducing the need for manual tagging and providing deeper insight into dynamic team behavior.
\end{abstract}


\section{Introduction}
\label{sec:intro}
Data analytics has become central to association football, spanning player recruitment, athletic performance, and tactical analysis \citep{rein_big_2016}. This scope has been further expanded through the introduction of tracking systems, recording the positions of all players and the ball at high frequency (10--25~Hz). Unlike traditional event-based data, tracking data captures all movements on- and off-ball, permitting a more comprehensive reconstruction of game situations and making tactical analysis an active area of research \citep{goes_unlocking_2021}.
Modeling this data while preserving the complete spatial arrangement of all players over time is a complex task. While several modeling approaches exist, no systematic comparison has identified a method that consistently outperforms all others across this complex setting in space and time. 
In particular, previous works have used graph neural networks \citep[GNNs; see][]{hamilton_graph_2020} that can handle the relational structure of football through permutation-invariant operations over player nodes \citep[see e.g.,][]{anzer_detection_2022, raabe_graph_2023, lee_contextual_2025, ochin_game_2025}. How GNNs compare to established tabular baselines such as gradient-boosted trees on tabular-based features, and how architectural choices such as graph topology and message-passing mechanism affect performance in a tactical setting, remain open questions which we address in this work. We study match phase identification as the evaluation task: it is tactically meaningful, as players' behavior, instructions, and positioning all depend on the current phase \citep[see e.g.,][]{shaw_dynamic_2019, ric_football_2021}, and tracking-synced labels are available at scale.

In this paper, we propose a GNN approach to capture spatial and temporal dependencies among football players, aiming to identify the current match phase for both teams at one-second resolution. Our contributions are as follows: 

First, our GNN uses player-level features directly, preserving the full spatial structure of all players instead of aggregating into team-wide features. We define a seven-class phase taxonomy that includes three defensive sub-phases (`High Block', `Mid Block', `Low Block'), distinguishable only by spatial player arrangements. Labels are generated by a rule-based model across 203 matches, enabling large-scale training and evaluation. \\
Second, we construct adjacencies using Delaunay triangulation, motivated by the triangular passing structures players form to keep passing options open. To capture the spatial and directional structure of these interactions, we adapt the edge-augmented graph transformer layer of \citet{shi_masked_2021} into a Spatial Edge-Augmented Convolution (SEAConv). Relational edge attributes (pairwise distance, relative angle, and team membership) enter both the attention weights and the messages themselves, and player coordinates are encoded with random Fourier features \citep{tancik_fourier_2020} to resolve fine-grained positional differences. We systematically evaluate both of these domain-driven design choices against common alternatives, as these properties have a direct influence on learning behavior in GNNs \citep{alon_bottleneck_2020, topping_understanding_2021, xu_how_2018, dwivedi_benchmarking_2023}. \\
Third, we make the model's predictions interpretable through the application of Integrated Gradients \citep[IG;][]{sundararajan_axiomatic_2017}, analyzing the importance of the graph's features, as well as the overall contribution of individual spatial features compared to graph-wide possession-based information.

Our work builds on two lines of research: tactical analysis from tracking data, and graph-based modeling in football. On the tactical side, tracking data has enabled applications such as team formation detection \citep{bauer_putting_2023}, pressing analysis \citep{lee_express_2025, bauer_data-driven_2021}, and play-style characterization \citep{shaw_dynamic_2019}.
Methodologically, GNNs have gained traction for modeling spatial tracking data. \citet{stockl_making_2021} show empirically that GNNs can make hand-crafted features unnecessary in certain football analytics tasks. \citet{raabe_graph_2023} propose `tactical graph networks' that transform tactical situations into graph representations to predict possession gain probability. \citet{ochin_game_2025} implement a graph-based architecture for event detection from tracking data. Graph-based representations of tracking data are not specific to football and have been applied across a range of invasion sports \citep[e.g.,][]{yeh_diverse_2019, sun_stochastic_2019, xenopoulos_graph_2021, raabe_graph_2023}.
Automated match phase detection has recently been approached from several directions. \cite{biermann_contextualization_2025} train a recurrent model on handcrafted team-level features for hierarchical binary classification of match phases with manually annotated labels. \cite{rothe2026tactical} define an expert state system that pairs in-possession and out-of-possession phases, including low, mid and high blocks, and detect these states by clustering the number of players per pitch zone within each ball zone, validated against human annotation of twelve Bundesliga matches.
Our work differs from these approaches in three respects. First, instead of zone-count representations, which discard inter-player distances and team compactness, we operate on player-level graphs that retain continuous positions and pairwise relations. Our taxonomy spans in-possession, out-of-possession, and transition phases, and the model itself must separate active phases from stoppages and unclassifiable play rather than relying on a pre-processing step. Second, we systematically compare adjacency constructions and message-passing layers across 80 runs, isolating how graph topology and message design affect tactical classification, a question that has received limited attention in football. Third, rule-based labels allow training and evaluation on 203 matches, substantially more than related work, at the cost of agreement with human annotation. 

The rest of the paper is structured as follows: Section~\ref{sec:graph_repr} introduces the data and graph representations. Section~\ref{sec:model} describes the message-passing layers and model architecture. Section~\ref{sec:emp_res} presents the empirical evaluation. In Section~\ref{sec:model_inter}, we interpret the model through graph embeddings and feature attributions. Section~\ref{sec:conclusion} discusses the results. 
All code is available at: \url{https://gitlab.kit.edu/vincent.renner/match_phase_detection}.

\section{Data and Graph Representations}
\label{sec:graph_repr}

\subsection{Data Set}
\label{sec:data_set}
The data set, provided by the Deutscher Fu{\ss}ball-Bund (DFB), comprises 203 matches from the DFB-Pokal, senior national team, and under-21 national team competitions. 
The data set covers men's competitions exclusively. Table \ref{tab:competition_season_descriptive} in the Appendix lists the number of matches per competition and season.
 
The data set comprises positional data at 25~Hz for all 22 players and the ball, described by spatial and kinematic features such as speed and acceleration. 
These data were collected by TRACAB\footnote{TRACAB provides an optical tracking system that originated within ChyronHego. TRACAB Technologies has operated as a standalone company since 2021 and was acquired by EA Sports in 2025.}, which has been used in various studies \citep[see e.g.,][]{bauer_putting_2023, bauer_data-driven_2021, anzer_expected_2022} and validated by \citet{linke_football-specific_2020, linke_validation_2018}. Event data were collected by Sportec Solutions. Both data sets were provided in the Common Data Format \citep{anzer_common_2025}.
The matches include the 2019/20--2024/25 seasons, and were selected based on data quality and the availability of match phase labels: for a total of six (sub)phases of play (`Build Up', `Low Block', `Mid Block', `High Block', `Counter Attack', `Defensive Transition'; see also Table \ref{tab:match_phases_descriptive}), we use rule-based labels (see Appendix \ref{app:rules}).
We construct a set of ten tailored spatial features outlined in Table \ref{tab:variables} in the Appendix.

\begin{table}[htbp]
	\caption{\textbf{Descriptive Statistics.} Share of phases in the data set and mean duration in seconds. Both are computed over the full data set of 203 matches. The numbers on the validation split differ slightly.}
	\label{tab:match_phases_descriptive}
	\centering
	\footnotesize
	\begin{tabular}{lccccccc}
		\toprule
		& \multicolumn{1}{c}{In Possession} 
		& \multicolumn{3}{c}{Out of Possession} 
		& \multicolumn{2}{c}{Transitions} 
		& \multicolumn{1}{c}{No Phase} \\
		\cmidrule(lr){2-2} \cmidrule(lr){3-5} \cmidrule(lr){6-7} \cmidrule(lr){8-8}
		& \makecell{Build Up}  
		& \makecell{High\\Block} 
		& \makecell{Mid\\Block} 
		& \makecell{Low\\Block} 
		& \makecell{Counter\\Attack} 
		& \makecell{Defensive\\Transition} 
		& \makecell{No Phase} \\
		\midrule
		Share         & 15.6\% & 0.9\% & 15.9\% & 2.1\% & 3.0\% & 5.3\% & 57.2\% \\
		Mean Duration & 16.6s  & 5.6s  & 13.9s  & 7.0s  & 6.3s  & 7.5s  & 22.7s  \\
		\bottomrule
	\end{tabular}
\end{table}
Every part of the match without an active phase for the observed team is labeled as `No Phase'. Keeping `No Phase' in our taxonomy is essential, as no prior information regarding start and end points of actual phases can be passed into the model in a real-world application. This results in seven (sub)classes within four larger categories as outlined in Table~\ref{tab:match_phases_descriptive}. The empirical distribution of the match phases and the mean duration of each is reported as well.
`No Phase' occupies the largest share of match time (57.2\%) as it consists of the union of stoppages and unclassifiable play, while `Build Up' and `Mid Block' are the most frequent labeled phases. 
Transition periods occupy less time, and `High Block', `Low Block', and `Counter Attack' are relatively rare (6.0\% in total).
Since both teams are labeled at each second, both teams' data can be used, effectively doubling the number of observations.
Further, we treat a full match as a single sequence from start to end. 

\subsection{Graph Representations}
\label{sec:model-adjacency-strategies}
\begin{figure}[ht]
	\centering
	\includegraphics[width=0.8\linewidth]{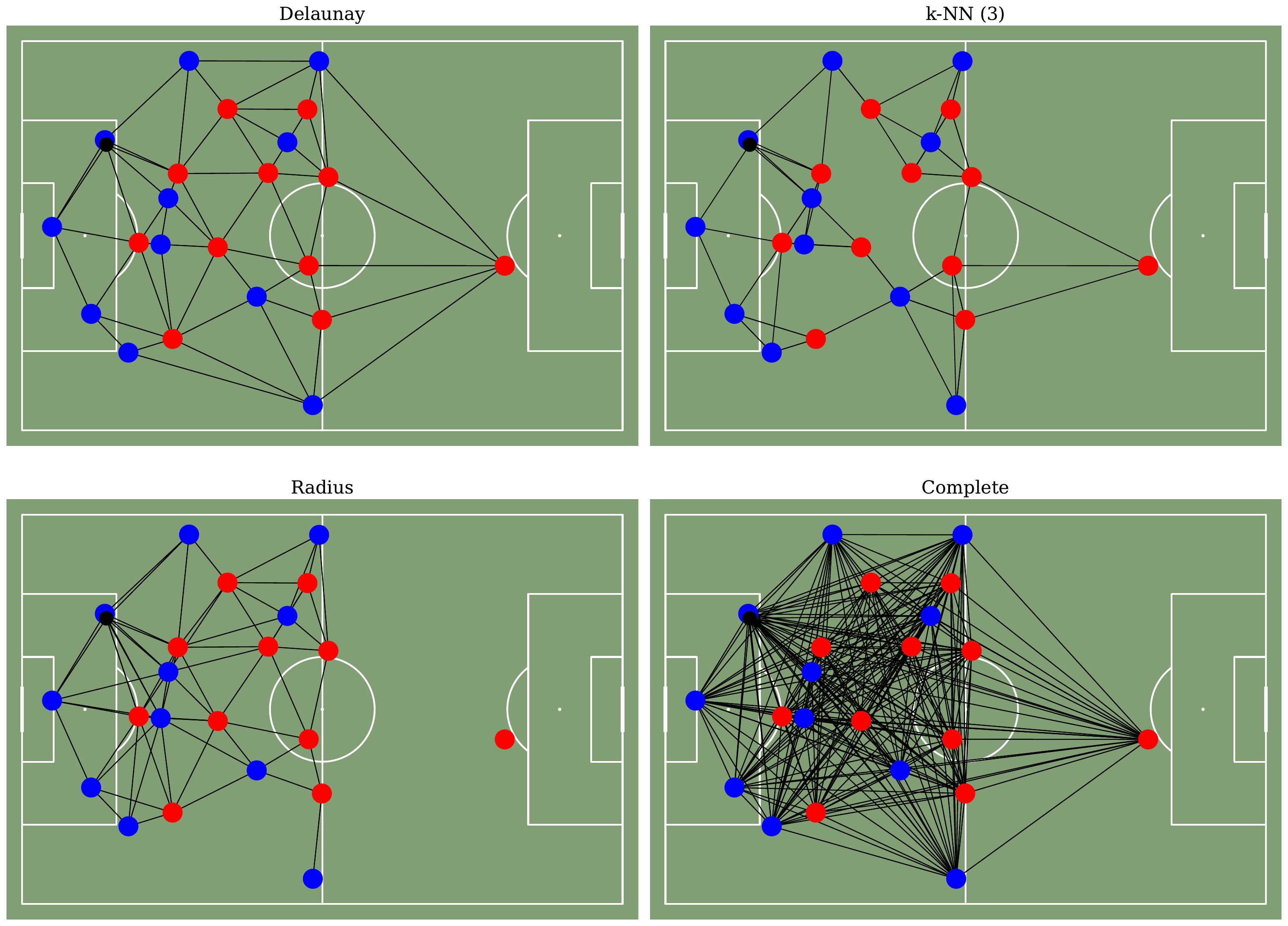}
	\caption{\textbf{Adjacency Construction Options.} Comparison of edge construction strategies for a single observation in the data set. Node colors indicate team associations, with the black node being the ball. \textit{Mexico vs. Germany, 02:26, International Friendly, October 17, 2023}.}
	\label{fig:edge_creation_comparison}
\end{figure}
We model each timestamp of the tracking data as a directed graph $G_t=(\mathcal{V}, \mathcal{E})$. $\mathcal{V}$ consists of all players and the ball, so $k=|\mathcal{V}| = 23$. Situations in which a player has been sent off are treated identically, with the corresponding reduction in the number of nodes. Each node $v_i$ carries an $m$-dimensional attribute vector $x_i \in \mathbb{R}^m$ comprising spatial coordinates (see Table~\ref{tab:variables}), movement descriptors, and indicators for the node type (attacking player, defending player, or ball) relative to the focal team, yielding a node feature matrix $X \in \mathbb{R}^{k \times m}$. All variables are listed in Table~\ref{tab:variables} in the Appendix. Connectivity is encoded by an adjacency matrix $A \in \{0,1\}^{k \times k}$ with $a_{i,j} = 1$ if edge $(v_i, v_j) \in \mathcal{E}$. Each edge carries an $l$-dimensional attribute vector $e_{i,j} \in \mathbb{R}^l$ containing the distance and angle between connected nodes as well as indicators of the relationship type (teammates, opponents, player-to-ball), collected in a feature tensor $E \in \mathbb{R}^{k \times k \times l}$. We set $e_{i,j} = \mathbf{0}$ if $a_{i,j} = 0$, where $\mathbf{0}$ denotes the zero-vector. We include graph-level features $u \in \{0,1\}^2$ indicating whether the focal team is in possession and whether the ball is in play. To reduce noise, one graph per second is created by averaging features across the 25 frames.

In other graph-based applications such as molecular systems \citep{qu_graph_2023, zhang_graph_2023}, edges are unique and observable, whereas edges in tracking data are not. Hence, the edge construction strategy poses a modeling decision, trading off computational cost against the quantity and type of information exchanged between nodes. Prior work typically uses complete graphs \citep[see e.g.,][]{wang_tacticai_2024, lee_contextual_2025} without systematic comparison of alternatives, leaving this design choice largely unexplored for football applications. \citet{raabe_graph_2023} reduce the complete graph based on different distance-based thresholds, showing that a sparse adjacency matrix can lead to improved results.
 
We propose to construct the adjacency matrix through Delaunay triangulation \citep{delaunay_sur_1934}, motivated by the spatial structure of football. Players typically arrange in triangles to create and maintain passing options; the Delaunay triangulation encodes this property by connecting nodes into triangles such that no other node lies inside the circumcircle of any triangle. Formally, given the set of spatial node coordinates
\[
	\mathcal{P}
	= \bigl\{\, x_i^{\mathrm{pos}} \in \mathbb{R}^2
	\mid i \in \mathcal{V} \bigr\},
\]
let $\mathcal{E}_{\mathrm{Del}}$ denote the edge set of the Delaunay triangulation $\mathcal{T}_{\mathrm{Del}}(\mathcal{P})$. The adjacency entry $a^{(\mathrm{Del})}_{i,j}$ is set to $1$ if $(i,j) \in \mathcal{E}_{\mathrm{Del}}$ and $0$ otherwise.
The Delaunay triangulation is dual to the Voronoi tessellation commonly used for pitch control analysis \citep[see e.g.,][]{taki_visualization_2000, chung_evaluation_2016}.
 
To evaluate this domain-informed construction, we compare against three general-purpose strategies. The k-nearest neighbors (k-NN) graph connects each node to its $k_{\mathrm{NN}}$ spatially closest nodes, using $k_{\mathrm{NN}}\in \{3,6\}$ in our evaluation setting. The radius graph connects all node pairs whose Euclidean distance falls below a threshold $r$, equivalent to the threshold-based sparsification used by \citet{raabe_graph_2023}. The complete graph connects every node to every other node, making no structural assumptions but incurring the highest computational cost, while being a common choice in prior work \citep[see e.g.,][]{wang_tacticai_2024, lee_contextual_2025}. In all constructions, self-loops are added to ensure each node's own attributes are processed during message passing. All adjacency matrices are symmetric except for the k-NN graph. Figure~\ref{fig:edge_creation_comparison} illustrates the four strategies applied to the same frame.
 
\section{Model Design}
\label{sec:model}

\subsection{GNN Layers}
\label{sec:model-gnn-layers}

Following \citet{gilmer_neural_2017}, a generic message-passing layer updates each node's feature vector as
\begin{equation*}
	x'_i = \gamma_{\Theta} \left( x_i, \bigoplus_{j \in \mathcal{N}(i)} \phi_{\Theta} (x_i, x_j, e_{j,i}) \right),
\end{equation*}
where $\gamma_{\Theta}$ and $\phi_{\Theta}$ are differentiable functions and $\bigoplus$ is a differentiable, permutation-invariant aggregation (e.g., sum, mean, or max), as generalized by \citet{fey_fast_2019}, and $\mathcal{N}(i)$ denotes the neighbors of node $i$. The message function $\phi_{\Theta}$ processes adjacent node features together with the connecting edge, and the update function $\gamma_{\Theta}$ combines the aggregated messages with the current node features.

Typically in GNNs, multiple message-passing layers are stacked, each receiving embeddings from its predecessor. In graph-level tasks, the final layer's representations are pooled through a readout layer 
\begin{equation*}
	z_G = R(\{x'_i \}),
\end{equation*}
which outputs a single vector for the entire graph. For the overall network to be invariant to graph isomorphism, the readout must depend only on the set of representations $\{x'_i\}$ and not on the node indexing \citep{gilmer_neural_2017}. This is realized through the same class of permutation-invariant operations used in neighborhood aggregation, ranging from sum, mean, or max pooling to learnable attention-based pooling \citep{li_gated_2015}.
 
In addition to standard layer choices, we propose a layer design adapted to match phase detection, addressing two limitations of existing operators. First, attention-based layers such as GATConv \citep{velickovic_graph_2017, brody_how_2022} incorporate edge attributes into the attention mechanism but not into the messages themselves. In our setting, edges encode distances, angles, and team membership --- information that should directly shape the messages exchanged between nodes. Second, raw low-dimensional spatial coordinates are subject to the spectral bias of neural networks toward low-frequency functions \citep{tancik_fourier_2020}, limiting the model's ability to distinguish nearby but tactically distinct positions. 

To address the first limitation, we adapt the edge-augmented graph transformer layer of \cite{shi_masked_2021} into a Spatial Edge-Augmented Convolution (SEAConv), a message-passing layer in which edge embeddings augment both the attention coefficients and the messages. Compared with \citet{shi_masked_2021}, SEAConv shares the key and value projection, treats the node's own state as a self-loop within the attention softmax instead of a separate skip connection, and operates on Fourier-encoded coordinates. The updated node features are computed as
\begin{equation*}
	x'_i = \frac{1}{K} \sum_{k=1}^{K}
	\sum_{j \in \mathcal{N}(i) \cup \{i\}}
	\alpha^{(k)}_{i,j} \left( h^{(k)}_{t,j}
	+ \tilde{e}^{(k)}_{i,j} \right),
\end{equation*}
where $h^{(k)}_{s,i} = \Theta_s^{(k)} x_i$ is the projection of the receiving node, which acts as the attention query, and $h^{(k)}_{t,j} = \Theta_t^{(k)} x_j$ is the projection of the sending neighbor. Augmented by the edge embedding, the latter serves as both key and value. The edge embeddings $\tilde{e}^{(k)}_{i,j} \in \mathbb{R}^{m'}$ are the $k$-th head component of a learnable edge encoder $\gamma_e: \mathbb{R}^l \to \mathbb{R}^{Km'}$ applied to the raw edge features $e_{i,j}$.
 
The attention coefficients follow scaled dot-product attention \citep{vaswani_attention_2017}:
\begin{equation*}
	\alpha^{(k)}_{i,j} = \mathrm{softmax}_j \!\left(
	\frac{\bigl\langle h^{(k)}_{s,i},\;
		h^{(k)}_{t,j} + \tilde{e}^{(k)}_{i,j} \bigr\rangle}
	{\sqrt{m'}}
	\right),
\end{equation*}
where $\mathrm{softmax}_j$ normalizes over $\mathcal{N}(i) \cup \{i\}$. The node's own state enters as a self-loop ($j=i$) whose raw edge features are set to $e_{i,i} = \mathbf{0}$. The $1/\sqrt{m'}$ scaling prevents softmax saturation as the head dimension grows. Attention weights are regularized via dropout \citep{velickovic_graph_2017} with probability $p_{\mathrm{attn}}$.
 
To address the second limitation, we encode the spatial coordinates $x_i^{\mathrm{pos}} \in \mathbb{R}^2$ via random Fourier features $\psi(\cdot) \in \mathbb{R}^{2m'}$, which map positions into a high-dimensional sinusoidal basis, enabling the network to resolve fine-grained spatial differences \citep{tancik_fourier_2020, rahimi_random_2007}:
\begin{equation*}
	\psi(x_i^{\mathrm{pos}}) = \left[
		\sin(B ^\top x_i^{\mathrm{pos}}),\;
		\cos(B^\top x_i^{\mathrm{pos}})
	\right],
\end{equation*}
where $B \in \mathbb{R}^{2 \times m'}$ is a fixed matrix with entries drawn from $\mathcal{N}(0, \sigma^2)$, and $\sigma$ controls the frequency band. The remaining non-spatial node attributes $x_i^{\mathrm{rest}} \in \mathbb{R}^{m - 2}$ are projected to $\mathbb{R}^{m'}$ via a separate linear layer $\Theta_r \in \mathbb{R}^{m' \times (m-2)}$ before concatenation, yielding $[\Theta_r x_i^{\mathrm{rest}} \;\|\; \psi(x_i^{\mathrm{pos}})] \in \mathbb{R}^{3m'}$.
As these modifications are directly targeted to enhance the handling of spatial attributes, subsequent stacked layers use standard graph attention layers (GATConv, introduced below), which in our implementation continue to consume the raw edge attributes but without the Fourier-based encodings.
 
We compare SEAConv against three established message-passing operators: Graph Convolutions \citep[GCNConv;][]{kipf_semi-supervised_2017}, which propagate features via the symmetrically normalized adjacency matrix; Graph Isomorphism Networks \citep[GINConv;][]{xu_how_2018}, which apply a learnable MLP after neighborhood aggregation; and Graph Attention Networks \citep[GATConv;][]{velickovic_graph_2017, brody_how_2022}, which weight neighbor contributions through multi-head attention. Formal definitions are provided in Appendix~\ref{app:baseline-layers}.

\subsection{Model Architecture}
Each graph is processed by a GNN block consisting of stacked message-passing layers followed by graph-level attention pooling \citep{li_gated_2015}. This transforms each graph into a single $d_g$-dimensional embedding $Z_t = \phi_g(G_t) \in \mathbb{R}^{d_g}$. We concatenate $Z_t$ with embedded global features $\phi_u(u_t)$, where $\phi_u$ is an MLP mapping the binary global features into a higher-dimensional representation. The resulting sequence is
\begin{equation}
	\label{eq:full-embedded-sequence}
	H_\text{seq} = ( Z_1 \| \phi_u(u_1),\;
	Z_2 \| \phi_u(u_2),\;
	\dots,\;
	Z_T \| \phi_u(u_T)).
\end{equation}

\begin{figure}[htbp]
	\centering
	\includegraphics[width=0.7\linewidth]{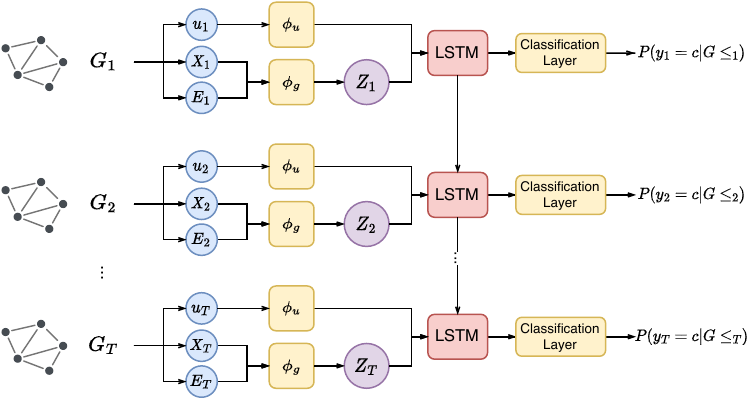}
	\caption{\textbf{Model Architecture}. Schematic structure of the
		model, receiving a sequence of graph representations as input and
	producing class probabilities for each timestep.}
	\label{fig:model-design}
\end{figure}

The concatenated representations are subsequently normalized using layer normalization and transformed by a learned projection before being passed to the recurrent model.
A unidirectional LSTM (\citealp{hochreiter_long_1997}; see Appendix~\ref{sec:lstm} for a short introduction) processes this sequence, producing an output state at each timestep. A final classification layer yields predicted probabilities
\begin{equation*}
	P(y_t = c \mid G_{\leq t}),
\end{equation*}
for each timestep $t$ and class $c$. The model architecture is visualized in Figure~\ref{fig:model-design}.
 
\subsection{Baseline Models}
\label{sec:baseline-models}

To isolate the contribution of graph representations, we compare against baselines operating on aggregated features.
We train four tabular baselines: multinomial logistic regression \citep[see e.g.,][]{hosmer_applied_2013} and XGBoost \citep{chen_xgboost_2016} on features consisting of the per-team mean and variance of the players' positions and velocities, as well as the global features; a separate XGBoost model using only the global features (possession and ball status); and an XGBoost model using pre-computed topological features derived from the point cloud of player positions. The construction of the topological features is explained in detail in Appendix~\ref{app:topo-features}. Since logistic regression and XGBoost cannot model temporal dependencies, each observation is augmented with lag features from the preceding ten seconds. 
The LSTM baseline shares the full model's sequence module and classification head, training protocol (data split, optimizer, learning-rate schedule, epochs) and seeds, and receives per-team aggregate features (centroid, positional variance and mean speed of both teams) in place of the pooled graph embedding $Z_t$, alongside the same global features. All models are trained and evaluated on the same match split.

\section{Empirical Results}
\label{sec:emp_res}

Our goal is to correctly classify match phases per second, where correctness is judged according to the given rule-based labels.
The GNN models are implemented in Python using PyTorch \citep{paszke_pytorch_2019} and PyTorch Geometric \citep{fey_fast_2019}. All models minimize multiclass cross-entropy (CE) loss (see Appendix~\ref{app:cross-entropy-loss}). 

Hyperparameters (Table~\ref{tab:hyperparameters}) were fixed before the comparative experiments and shared across all adjacency–layer combinations and seeds. Where established values exist, we adopted them from the literature (optimizer settings, weight decay, GNN and attention dropout, hidden dimension, number of message-passing layers; sources in Table \ref{tab:hyperparameters}). The remaining settings (the Fourier scale $\sigma$, the radius threshold r, the number of attention heads, the learning-rate floor, the LSTM dropout, and the number of LSTM layers) were set manually in preliminary runs with the sole aim of stable convergence. No systematic hyperparameter search was performed. This is deliberate, as the purpose of this study is not to maximize predictive performance, but to compare graph constructions and message-passing operators under identical training conditions. Additionally, its aim is to assess whether a graph-based architecture can learn tactical phase structure from player-level tracking data. Because all configurations share the same hyperparameters and training protocol, performance differences between them can be attributed to the design choice under study rather than to differential tuning effort. Tuning each configuration separately would likely raise absolute performance, but would confound this comparison.

The data set is split 70/30 into training and validation sets by full matches to avoid data leakage. We do not hold out a separate test set. The validation set is used only for reporting, and configuration-level results are means over all runs of that configuration. Absolute performance figures may therefore be slightly optimistic as estimates of performance on unseen matches. The same protocol applies to every model, so this does not systematically favor any configuration in the relative comparisons, which are the focus of this work. All GNN results in Tables \ref{tab:design_choice_results} and \ref{tab:baseline_results} are reported as means over four seeds. 

To accommodate over 5,000 observations per team and match, truncated backpropagation through time (BPTT) is applied, with each sequence split into four equal parts. Before the actual training, the graph encoder (GNN layers, attention pooling, and global-feature embedding) is pre-trained for 10 epochs through an auxiliary frame-wise linear classification head, bypassing the LSTM. The main training loop then starts after this pre-training stage.

We evaluate validation performance using accuracy, balanced accuracy $\mathrm{BAcc}=\frac{1}{C} \sum_{c=1}^{C} \frac{\text{TP}_c}{\text{TP}_c + \text{FN}_c}$, and macro F1 score $
\mathrm{F1}_{\mathrm{macro}} = \frac{1}{C} \sum_{c=1}^{C} \frac{2\,\text{TP}_c}{2\,\text{TP}_c + \text{FP}_c + \text{FN}_c}$ with $C$ the number of classes. Note that `raw' accuracy can be misleading given the class imbalance present in our data set: `No Phase' accounts for 56.3\% of observations. Therefore a trivial classifier predicting the majority class would already achieve 56.3\% accuracy. Balanced accuracy and macro F1 are therefore our primary evaluation metrics.
Metrics are reported using the mean of the last 10 epochs.

\subsection{Graph Representations and Message-Passing Layers}
\label{sec:graph-rep-results}
    
\begin{table}[htbp]
	\centering
	\caption{\textbf{Design Choice Results.} Mean performance metrics for adjacency-construction strategies (Panel A) and message-passing layers (Panel B), averaged over all seeds and combinations of other parameters. The best metric per panel is in bold; relative differences are shown in parentheses. Higher values indicate better performance for all metrics except CE loss, where lower is better.}
	\label{tab:design_choice_results}
	\begin{tabular}{l@{\hskip 0.2in}ccccc}
		\toprule
		                  & CE Loss         & Accuracy       & Bal. Accuracy  & Weighted F1    & Macro F1       \\
		\midrule
		\multicolumn{6}{l}{\textit{Panel A: Adjacency Construction}} \\
		\midrule
		Delaunay          & \textbf{0.249}  & \textbf{0.901} & \textbf{0.773} & \textbf{0.898} & \textbf{0.788} \\
		k-NN ($k_{NN}=3$) & 0.254 (+2.0\%)  & 0.899 (-0.2\%) & 0.767 (-0.8\%) & 0.897 (-0.2\%) & 0.783 (-0.6\%) \\
		Radius            & 0.258 (+3.8\%)  & 0.899 (-0.3\%) & 0.767 (-0.8\%) & 0.896 (-0.3\%) & 0.783 (-0.7\%) \\
		k-NN ($k_{NN}=6$) & 0.258 (+3.7\%)  & 0.898 (-0.3\%) & 0.761 (-1.5\%) & 0.895 (-0.3\%) & 0.780 (-1.0\%) \\
		Complete          & 0.276 (+11.2\%) & 0.894 (-0.8\%) & 0.749 (-3.1\%) & 0.891 (-0.9\%) & 0.767 (-2.6\%) \\
		\midrule
		\multicolumn{6}{l}{\textit{Panel B: Message-Passing Layer}} \\
		\midrule
		SEAConv           & 0.248 (+0.0\%)  & \textbf{0.902} & \textbf{0.781} & \textbf{0.899} & \textbf{0.793} \\
		GATConv           & \textbf{0.247}  & 0.902 (-0.0\%) & 0.775 (-0.8\%) & 0.899 (-0.0\%) & 0.788 (-0.6\%) \\
		GCNConv           & 0.258 (+4.1\%)  & 0.897 (-0.5\%) & 0.762 (-2.5\%) & 0.894 (-0.6\%) & 0.778 (-2.0\%) \\
		GINConv           & 0.283 (+14.4\%) & 0.893 (-1.0\%) & 0.737 (-5.7\%) & 0.889 (-1.1\%) & 0.761 (-4.1\%) \\
		\bottomrule
	\end{tabular}
\end{table}

We evaluate our two proposed modeling choices, Delaunay triangulation for adjacency construction and SEAConv for message passing, against the alternatives introduced in Sections~\ref{sec:model-adjacency-strategies} and~\ref{sec:model-gnn-layers}. We fit models for each combination of adjacency strategy and message-passing layer with four different seeds, yielding 80 runs. To isolate the effect of each design choice, we aggregate metrics across all combinations of the remaining parameters.

Table~\ref{tab:design_choice_results} shows the results for both design choices. Regarding the adjacency construction (Panel A), the differences among sparse strategies are modest, but the ranking is mostly consistent across metrics. Delaunay triangulation performs best, followed by the k-NN ($k_{\mathrm{NN}}=3$) graph. The complete graph ranks last. 
To verify these rankings statistically, we conduct pairwise Wilcoxon signed-rank tests on the Macro F1, pairing runs by the complementary factor (message-passing layer) and seed. Full results are reported in Table~\ref{tab:wilcoxon_tests} in the Appendix. The block-normalized residuals in Figure~\ref{fig:experiment_results_residuals} provide a descriptive view of robustness that is quantified by the Wilcoxon analysis. Delaunay scores significantly higher than every alternative (Holm-adjusted $p \leq 0.0003$ in all four comparisons; higher Macro F1 in at least 15/16 matched pairs in each case), with bootstrap 95\% CIs on the mean paired difference excluding zero. The only non-significant pairwise contrast is $k_{\mathrm{NN}}=3$ vs. Radius ($p = 0.50$), consistent with their near-identical aggregate metrics.
The superiority of the Delaunay graph is an interesting result, as the complete graph is a common choice in prior work \citep[see e.g.,][]{wang_tacticai_2024, lee_contextual_2025}. A likely explanation for the difference in performance is that sparser adjacency matrices prune edges between distant players that carry little information, while relationships between non-adjacent players can still be captured through multi-hop message passing. Delaunay graphs may exceed other sparse adjacency matrices by apparently keeping more meaningful edges and maintaining the spatial structure of the graph. The complete graph also incurs substantially higher computational cost: it contains roughly $4.5\times$ more edges than a Delaunay graph, leading to a 67\% increase in training time on average in our experiments.
 
Having established the advantage of Delaunay adjacency, we turn our attention toward the message-passing layers (see Panel B of Table~\ref{tab:design_choice_results}). SEAConv achieves the best performance on every metric except the CE loss, where the difference to the GATConv is marginal. The improvements are modest but consistent across all runs and adjacency strategies, suggesting that adapting the message-passing function to the problem structure yields reliable, if incremental, gains over standard layers.
The same pairwise Wilcoxon analysis (Panel B of Table~\ref{tab:wilcoxon_tests}) confirms that SEAConv significantly outperforms every alternative on Macro F1, including the close competitor GATConv (mean paired difference $+0.0050$, 95\% CI $[+0.0042, +0.0060]$, $p < 10^{-4}$ Holm-adjusted, higher in 20/20 matched pairs). The margin over GATConv is small in absolute terms, but it is statistically reliable across the full crossing of adjacency strategies and seeds.

As the Delaunay adjacency combined with SEAConv led to the best performance, we use this configuration for all subsequent analyses and refer to it as `our' model. Full results for all 80 runs are reported in Table~\ref{tab:full_results} in the Appendix.

\subsection{Comparison with Baselines}
\label{sec:baseline-models-results}
We now compare our GNN configuration against the baseline models to assess whether a graph-based approach justifies its added complexity. All models are trained on the same matches and evaluated on the same validation matches.
Table~\ref{tab:baseline_results} reports the results. The GNN outperforms every baseline on every metric. The strongest baseline is XGBoost on the lagged standard features, which trails the GNN by only 2.4\% in accuracy but by 3.7\% in balanced accuracy, 4.6\% in macro F1 and 28.8\% in CE. 
The gap is thus small in accuracy but substantial in cross-entropy and on the less frequent phases, where the class-balanced metrics diverge. The LSTM, which shares the GNN's training protocol and classification head, follows closely in CE (+29.7\%) but falls further behind on macro F1 (-8.4\%). 
Each of the four GNN seeds individually outperforms both the XGBoost and the LSTM baseline on every metric. Hand-crafted interaction features (convex-hull areas, persistence entropies) do not close it, and perform well below the standard features. This indicates that a learned representation of player interactions captures information that neither aggregate positional features nor hand-crafted spatial descriptors provide.

\begin{table}[htbp]
	\centering
	\caption{\textbf{Baseline Results.} Comparison of the GNN with the baseline models across evaluated metrics on the validation set. Percentages are relative to the GNN. The GNN metrics refer to the mean across seeds of the model using Delaunay adjacency and SEAConv layers (see Table \ref{tab:full_results}). XGBoost is abbreviated as `XGB'. Percentages are computed from unrounded values.}
	\label{tab:baseline_results}
	\begin{tabular}{l@{\hskip 0.2in}ccccc}
		\toprule
		Model              & CE Loss          & Accuracy       & Bal.\ Accuracy  & Weighted F1     & Macro F1        \\
		\midrule
		GNN (ours)         & \textbf{0.241}   & \textbf{0.904} & \textbf{0.785}  & \textbf{0.901}  & \textbf{0.797}  \\
		XGB -- Standard    & 0.310 (+28.8\%)  & 0.882 (-2.4\%) & 0.756 (-3.7\%)  & 0.878 (-2.6\%)  & 0.761 (-4.6\%)  \\
		LSTM               & 0.313 (+29.7\%)  & 0.876 (-3.0\%) & 0.749 (-4.6\%)  & 0.875 (-2.9\%)  & 0.731 (-8.4\%)  \\
		XGB -- Topological & 0.366 (+51.9\%)  & 0.856 (-5.3\%) & 0.614 (-21.8\%) & 0.847 (-6.0\%)  & 0.625 (-21.6\%) \\
		Multinomial Logit  & 0.447 (+85.3\%)  & 0.856 (-5.3\%) & 0.660 (-16.0\%) & 0.849 (-5.7\%)  & 0.687 (-13.9\%) \\
		XGB -- Globals     & 0.519 (+115.3\%) & 0.822 (-9.0\%) & 0.469 (-40.2\%) & 0.799 (-11.3\%) & 0.446 (-44.0\%) \\
		\bottomrule
	\end{tabular}
\end{table}
\subsection{Calibration}

\begin{figure}[htbp]
	\centering
	\includegraphics[width=0.5\linewidth]{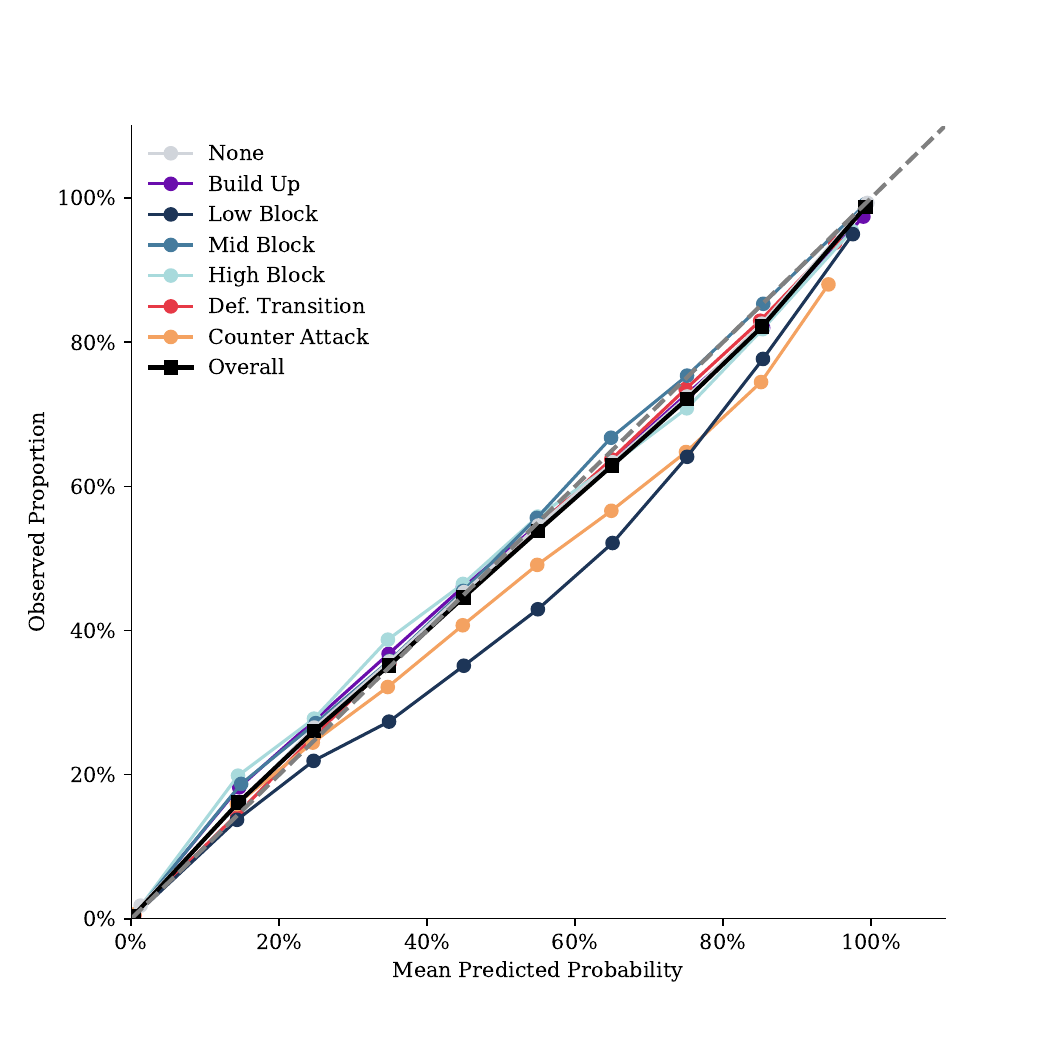}
	\caption{\textbf{Reliability Diagram.} Per-class calibration of predicted probabilities from the GNN model (all four Delaunay / SEAConv runs on the last epoch), with mean predicted probability plotted against empirical class frequency per confidence bin. }
	\label{fig:calibration-plot}
\end{figure}

In the preceding section, we evaluated the model's ability to discriminate between classes.
We now assess the quality of the model's probabilistic predictions. Formal definitions are provided in Appendix~\ref{app:calibration-definitions}.
 
Figure~\ref{fig:calibration-plot} visualizes the per-class reliability diagram. For most classes, the calibration curves lie close to the diagonal, indicating good calibration across both majority and minority classes. The `Low Block' and `Counter Attack' classes are systematically overconfident in the 0.3-0.8 range.
 
To quantify calibration error, we use the expected calibration error (ECE), which measures the weighted average absolute difference between observed positive fractions and mean predicted probabilities across bins \citep{pakdaman_naeini_obtaining_2015}. Table~\ref{tab:calibration_metrics} reports the ECE alongside per-class CE and F1 score. Given the true class, CE is highest for the transition phases and `High Block'. Consequently, the F1 score also drops for these classes. In settings with high class imbalance, this can be expected. The overall CE hides this because frequent classes dominate it. The model achieves low ECE across all classes.

\section{Model Interpretability}
\label{sec:model_inter}

Beyond predictive performance, we investigate whether the graph-based model captures the football-specific structure in the tracking data. We use Integrated Gradients \citep[IG;][]{sundararajan_axiomatic_2017} to attribute predictions to inputs. For a model $F: \mathbb{R}^n \to [0,1]$, input $z \in \mathbb{R}^n$, and baseline $z' \in \mathbb{R}^n$, IG computes the path integral of gradients from $z'$ to $z$:
\begin{align}
	\text{IG}_i(z) & \coloneqq \, (z_i - z_i') \cdot                                                                                              
	\int_{\alpha=0}^{1} \frac{\partial F(z' + \alpha \cdot (z - z'))}{\partial z_i} \,\mathrm{d}\alpha \nonumber\\ 
	               & \approx \, (z_i - z_i') \cdot \sum_{s=1}^{S}\frac{\partial F(z' + \frac{s}{S} \cdot (z-z'))}{\partial z_i} \cdot \frac{1}{S} 
	\label{eq:integ-grads-approx}.
\end{align}
For our multiclass model, we apply IG separately to each class probability $F_c : \mathbb{R}^{n} \rightarrow[0,1]$, yielding per-class attributions $\text{IG}_i^{(c)}(z)$ for $c=1,\dots,C$. Attributions for class $c$ are aggregated over all observations annotated with label $c$, so that the reported values indicate the model's use of each input while phase $c$ is active. 
We apply IG in two analyses: first to assess the contribution of graph embeddings versus global features (Section \ref{sec:importance-graph-embeddings}), second to identify the most influential node features (Section \ref{sec:relevance-node-features}).

\subsection{Importance of Graph Embeddings}
\label{sec:importance-graph-embeddings}

As the baseline LSTM already achieves reasonable performance using only information about possession, ball-in-play indicators, and aggregated positional data (refer to Table~\ref{tab:baseline_results}), we examine whether the pooled graph embeddings $\phi_g(G_t)$ provide additional predictive information beyond the embedded global features $\phi_u(u_t)$.
We compute IG on the full embedded sequence with $z =H_{\text{seq}}$ (Equation~\ref{eq:full-embedded-sequence}) using the validation set mean embedding as baseline $z'$. All analyses were carried out on the validation set exclusively.
 
\begin{figure}[htbp]
	\centering
	\includegraphics[width=0.6\linewidth]{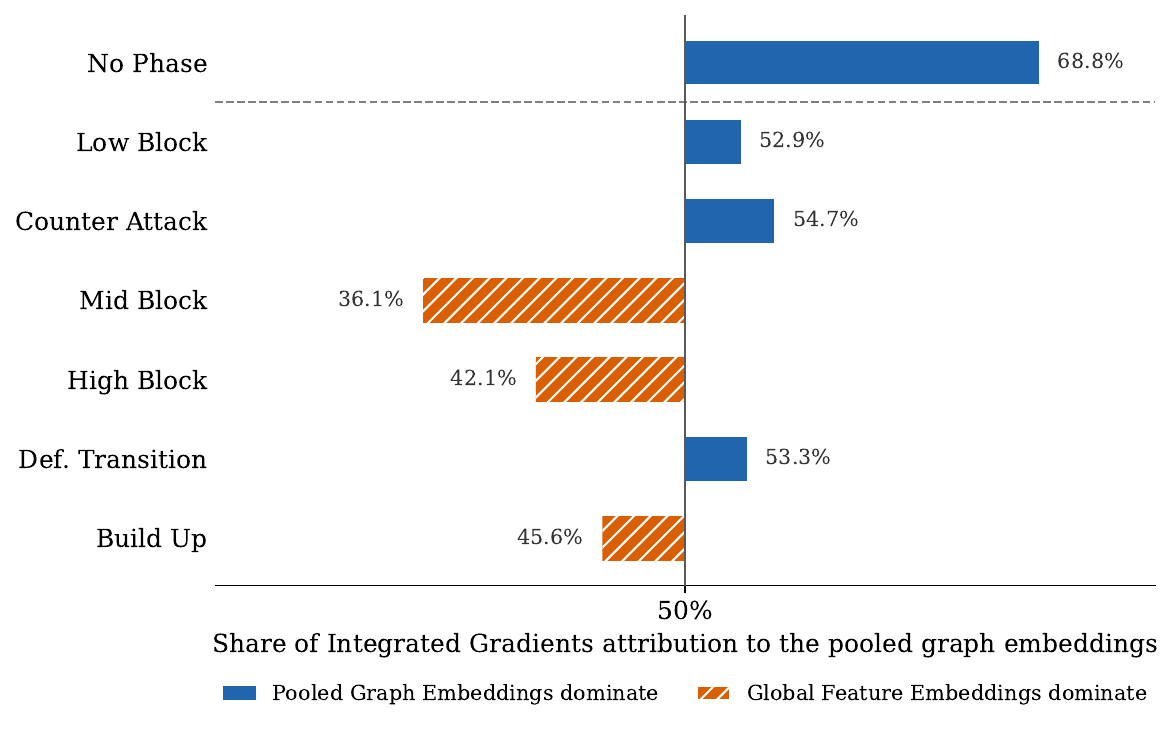}
	\caption{\textbf{Importance of Embeddings.} 
		Per-class share of total Integrated Gradients importance $\phi_g(G_t) +\phi_u(u_t)$ attributed to the pooled graph embeddings $\phi_g(G_t)$ (computed via Equation~\ref{eq:integ-grads-approx}).}
	\label{fig:importance-graph-embeddings}
\end{figure}
 
Figure~\ref{fig:importance-graph-embeddings} shows that the model relies on both input types across all classes, with the graph embeddings receiving at least 36.1\% of the overall attribution, while reaching higher attributions than global features in the `Low Block', `Counter Attack', `Def. Transition', and `No Phase' phases. The global features take the larger share in `Build Up', `Mid Block', and `High Block'. Every phase except `No Phase' occurs under one dominant combination of possession and ball-in-play state (Panel~B of Table~\ref{tab:global-state-distribution}). Hence, $\phi_u(u_t)$ alone already separates attacking from defending phases. The one exception is the `No Phase' class: as the only phase spread across all four global states, it is also the phase for which the graph embeddings receive their largest share (68.8\%). Within a global state, by contrast, the individual phases cannot be resolved. `Low Block', `Mid Block', and `High Block' share the same dominant state in 95.3\%, 97.4\% and 98.1\% of their frames, respectively. Conditioning on that state leaves the phase undetermined among them (Panel~A: `Mid Block' 53.1\%, `Def.\ Transition' 14.2\%, `Low Block' 6.6\%, `High Block' 3.2\%). Separating these phases requires information about the spatial configuration. This result is consistent with Section~\ref{sec:baseline-models-results}, where the LSTM with access to the same global features but aggregated spatial information is surpassed by the full model.
This comparison comes with one caveat. As IG attributions scale with the displacement $z - z'$ from the baseline (Equation~\ref{eq:integ-grads-approx}), for which we use the validation set mean embedding, aggregating over the frames of a given phase favors whichever stream is most strongly displaced from its marginal mean during that phase. Therefore, Figure~\ref{fig:importance-graph-embeddings} reports relative reliance under this baseline rather than the unique information each stream carries.

\begin{figure}[htbp]
	\centering
	\includegraphics[width=0.75\linewidth]{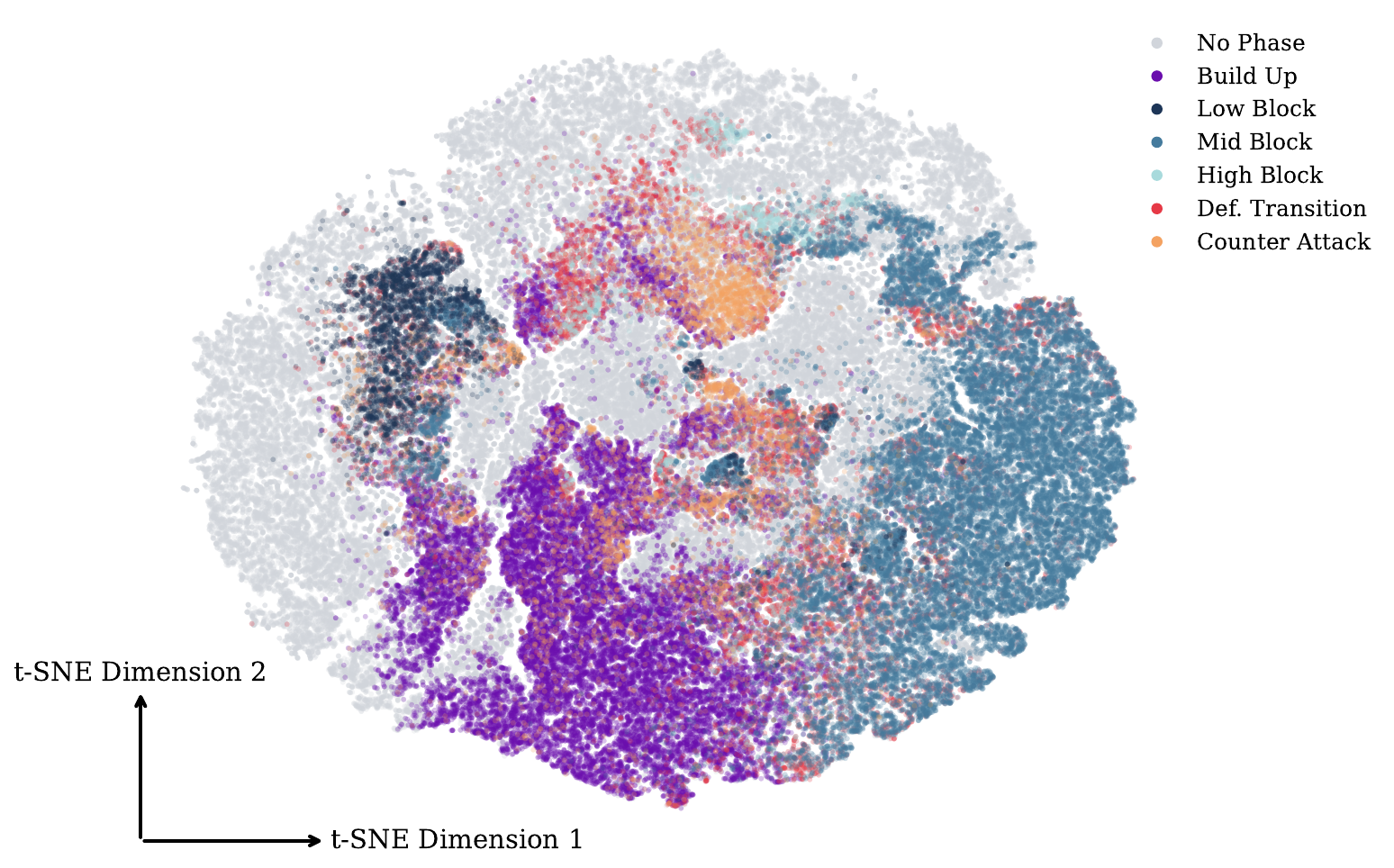}
	\caption{\textbf{t-SNE Embeddings.} Learned graph embeddings per class reduced to two dimensions using t-SNE. A random 25\% subsample of the validation set is shown for readability.}
	\label{fig:tsne-embeddings}
\end{figure}
 
To assess the learned representations visually, we project the 64-dimensional embeddings $\phi_g(G_t)$ into two dimensions using t-SNE \citep{van_der_maaten_viualizing_2008} and visualize them in Figure~\ref{fig:tsne-embeddings}. Classes appear well separated in this projection, indicating that the embeddings capture phase-specific structure.
 
\subsection{Relevance of Node Features}
\label{sec:relevance-node-features}
We further apply IG to quantify the importance of individual node features. To isolate their contribution, we compute IG with $z=\operatorname{vec}(X)$, the flattened node-feature matrix, holding adjacency, edges, and globals fixed at their original values. The baseline $z'$ is set to the unconditional feature means across all nodes in the validation set. Per-class attributions are computed for every validation set observation, reduced to absolute values, mean-pooled across all nodes of each graph, and then averaged over all observations annotated with phase $c$ to obtain a single attribution per combination of input feature and class. For concepts represented by multiple features (e.g., team association), we use the sum as aggregation. 

\begin{table}[htbp]
	\centering
	\caption{\textbf{Feature Importance.} Relative node feature importance by class per Integrated Gradients (Equation~\ref{eq:integ-grads-approx}). Variables are x (X) and y (Y) coordinates, speed (S), acceleration (A), distance to opponent goal (DG), distance to ball (DB), team association (TA; sum of three variable contributions), movement direction (MD; sin + cos), angle to goal (GA; sin + cos), and angle to ball (BA; sin + cos).}
	\label{tab:node_feature_importance_by_class}
	\begin{tabular}{l@{\hskip 0.2in}cccccccccc}
		\toprule
		                & X      & Y     & S      & A     & DG    & DB    & TA     & MD    & GA    & BA     \\
		\midrule
		No Phase        & 26.1\% & 7.7\% & 12.9\% & 3.4\% & 0.7\% & 7.0\% & 19.3\% & 8.6\% & 2.2\% & 12.2\% \\
		Build Up        & 18.0\% & 6.7\% & 10.0\% & 6.4\% & 0.5\% & 8.0\% & 23.9\% & 7.3\% & 1.9\% & 17.4\% \\
		Low Block       & 36.3\% & 5.1\% & 7.8\%  & 5.4\% & 0.8\% & 6.5\% & 20.2\% & 6.0\% & 1.0\% & 10.8\% \\
		Mid Block       & 16.9\% & 8.1\% & 11.3\% & 6.6\% & 0.6\% & 7.3\% & 24.6\% & 9.9\% & 1.2\% & 13.5\% \\
		High Block      & 40.1\% & 5.4\% & 7.0\%  & 4.4\% & 0.7\% & 5.0\% & 17.5\% & 8.5\% & 1.7\% & 9.6\%  \\
		Def. Transition & 22.3\% & 6.3\% & 10.5\% & 6.5\% & 0.7\% & 7.1\% & 22.1\% & 9.6\% & 1.5\% & 13.6\% \\
		Counter Attack  & 18.4\% & 6.3\% & 10.2\% & 6.4\% & 0.7\% & 6.8\% & 25.5\% & 9.6\% & 1.5\% & 14.5\% \\
		\bottomrule
	\end{tabular}
\end{table}
 
Table~\ref{tab:node_feature_importance_by_class} reports the relative importance of each node feature by output class. Values are normalized within each class to sum to 100\%. The x-coordinate is the most important feature in four classes, with its highest attributions for `High Block' (40.1\%) and `Low Block' (36.3\%). This aligns with the fact that horizontal position directly distinguishes defensive phases at different pitch depths. The x-coordinate outweighs the y-coordinate in all classes, reflecting the stronger role of horizontal positioning in phase differentiation. 
Team association is most important in the other three classes (23.9--25.5\%), indicating that the one-hot encoding of team membership is effectively learned. However, it must be stated that this feature aggregates three one-hot columns, with the per-column contribution of roughly 6--9\% remaining below the x-coordinate's in every class. 
Among dynamic features, angle to ball is the largest contribution in all classes but `No Phase' (9.6--17.4\%), with speed (7.0--12.9\%) and movement direction (6.0--9.9\%) showing a moderate contribution. Distance to goal shows a very low contribution, which can be explained through the feature not containing much additional information beyond the x-coordinate.

Together, these analyses indicate that the model not only performs well (see Section~\ref{sec:emp_res}) but also bases its predictions on tactically plausible features, consistent with domain knowledge of defensive phases.

\section{Discussion} 
\label{sec:conclusion}
The central finding of this paper is that spatial relationships between players carry predictive information for match phase classification that tabular feature representations do not capture. Our GNN-LSTM architecture outperformed all tabular baselines, whilst the strongest, XGBoost on the standard features, scores 4.6\% lower in macro F1 and incurs 28.8\% higher cross-entropy. This suggests that the processing of the full spatial configuration of the players is advantageous for detecting tactical patterns compared to aggregated features present in the baselines. Additionally, the interpretability analysis reinforces this, assigning comparable importance to the graph embeddings and the global possession state features.

Two design choices contributed to this result. Delaunay triangulation outperformed all alternative adjacency strategies, while complete graphs ranked last with 11.2\% higher cross-entropy. Removing edges between distant players appears to reduce uninformative messages during aggregation, while multi-hop passing still allows information flow across the pitch. The computational savings are a positive side effect: complete graphs required 67\% longer training time. Our custom SEAConv layer, which incorporates edge attributes directly into messages rather than only into attention weights, ranked first among the message-passing operators tested. The margins over GCNConv and GINConv are substantial (2.0\% and 4.1\% lower macro F1), whereas the margin over the closest competitor, GATConv, is small in absolute terms (tied in cross-entropy, +0.005 macro F1) though statistically reliable across all runs and adjacency strategies. Identifying which specific design elements within SEAConv drive this advantage, and whether its edge over attention-based message passing holds on other data sets, remains an open question for future work.
 
The IG analysis provides further support for the graph-based approach. Graph embeddings received at least 36.1\% of the total attribution across all classes, and held the majority share in four of seven. The global features alone cannot separate the three block phases, which occur almost exclusively under the same global state (ball in play, opponent in possession), so the model must draw on spatial player configurations rather than game-state indicators alone. At the node-feature level, the x-coordinate was the most important input feature next to the team association, which makes tactical sense: classifying defensive phases depends on where defensive lines sit along the length of the pitch. The reliability diagram (see Figure~\ref{fig:calibration-plot}) indicates that predicted class probabilities are well calibrated for all classes except `Low Block'.

The proposed framework offers practical relevance, as it automates phase labeling, a task currently performed manually. 
As the architecture is applicable beyond the specific phase taxonomy used in this work, it can be retrained on the custom phase definitions commonly used by professional teams, widening its applicability.

The practical value of the graph-based representation extends beyond this specific task. Because a match is encoded as relations between individual nodes, the same representation supports the detection of tactical patterns at the individual, group, and team level within a single framework. Previous approaches are typically tied to one of these levels or to one specific pattern: image-based representations capture team-level structures such as the basic phases of play considered in this study \citep{bauer_putting_2023}, whereas the detection of dedicated patterns --- for instance counter-pressing \citep{bauer_data-driven_2021} or counter attacks \citep{fassmeyer_toward_2021} --- has so far relied on pattern-specific, hand-crafted feature representations. Combinations of image-based and LSTM representations have been used to classify the defensive roles of individual players at corner kicks \citep{bauer_individual_2022}. Graph representations, providing a task-agnostic, permutation-invariant representation of the spatio-temporal multi-agent data, have already been successfully used in sports, e.g., to classify group-level patterns like overlapping runs \citep{anzer_detection_2022}.

Beyond pattern detection, an efficient graph representation of match situations can serve as a shared backbone for more sophisticated downstream models. While movement prediction with graph neural networks has already been demonstrated \citep{omidshafiei_multiagent_2022}, expected possession value (EPV) models represent a further, not yet fully exploited potential of graph-based representations.

Our proposed approach comes with several limitations. First, performance dropped for transition classes `Counter Attack' (F1 = 0.54) and `Defensive Transition' (F1 = 0.66). These phases are typically very short, so their boundaries fall within a few seconds and a minor delay in detection costs a large share of the class (see Table~\ref{tab:match_phases_descriptive}). Using the full 25 Hz frame rate rather than aggregating to one-second graphs could enhance the model for these classes, but would increase computational cost. Second, our labels are generated by a rule-based system rather than manual annotation, meaning label quality depends on the fidelity of that system. Third, the data set covers only men's football and a limited number of competitions, and generalization to women's football data or to a larger, more diverse set of competitions remains untested. Fourth, no independent test set was used, and hyperparameters were not tuned per configuration. Our results should therefore be read as a controlled comparison of design choices under a fixed training protocol rather than as optimized performance estimates. Confirming the absolute performance on an independent set of matches, ideally from further competitions, is left to future work.

Beyond these limitations, our findings suggest that domain-informed graph construction and edge-aware message passing may benefit other tactical tasks operating on tracking data, motivating future work on graph-based modeling across a wider range of spatio-temporal sports analytics.
 
\cleardoublepage\phantomsection\addcontentsline{toc}{section}{Bibliography}
\label{bib}
\bibliography{references}

\clearpage

\section*{Appendix}
\addcontentsline{toc}{section}{Appendix}

\appendix

\section{Definition of Play Phases}
\label{app:rules}
This appendix formalizes the rule-based labels of the six phases of play reported in this work --- `Build Up', `Low Block', `Mid Block', `High Block', `Counter Attack', `Defensive Transition' --- and extends the five-phase scheme of \citet{bauer_putting_2023}. A phase is kept alive until the ball goes dead. Distances are given in meters, speeds in km/h, and durations in seconds or milliseconds. The pitch is partitioned into an own third, a middle third, and a final third, referenced from the perspective of the team the phase describes, with the final third adjacent to the opponent's goal.

\subsection{General Concepts}

\paragraph{Stationary} A player is considered stationary if it is not moving forward or backward at more than $10$\,km/h, where forward and backward movement are defined by two opposing $60^\circ$ cones aligned with the goal-to-goal axis.

\paragraph{Organized} A team is considered organized if its units approximately form horizontal lines. Excluding the most advanced player, the remaining defenders are grouped into three units; letting $a$, $b$, and $c$ denote the vertical band width of these units from rear to front, the organization score is $(4a + 4b + c)/9$, and the team is organized if this score is below $250$. 

\subsection{Build Up}
`Build Up' is an in-possession phase that requires at least $8$ defenders behind the ball, independent of the third in which the ball is located. A build up is classified as opposed if the defending team is pressing and the closest defender is within $12$\,m of the ball, or at least two defenders are within $6$\,m of the ball, or at least one player is within $4$\,m of the ball. The phase is kept alive until its criteria are not fulfilled for $500$\,ms within a $2.5$\,s window. 
\subsection{Low Block}
`Low Block' is an out-of-possession phase that requires the defending team to be stationary and organized. The team has been stationary for at least $1.6$\,s during the previous $2$\,s; at least $4$ defenders are behind the ball and organized, or at least $4$ defenders are in the own box; at least $5$ defenders are in the own third; and at least $8$ defenders are in the middle or own third. The phase is kept alive until its criteria are not fulfilled for $400$\,ms within a $2$\,s window.

\subsection{Mid Block}
The team has been stationary for at least $1.6$\,s during the previous $2$\,s; at least $4$ defenders are behind the ball and organized, or at least $4$ defenders are in the own box; no low block applies; and at least $8$ defenders are in the middle or own third. The phase is kept alive until its criteria are not fulfilled for $400$\,ms within a $2$\,s window.

\subsection{High Block}
The team has been stationary for at least $1.6$\,s during the previous $2$\,s; at least $4$ defenders are behind the ball and organized; no mid or low block applies; at least $3$ defenders are in the final third; at least $8$ defenders are in the middle or final third; and possession changed more than $500$\,ms ago. The phase is kept alive until its criteria are not fulfilled for $400$\,ms within a $2$\,s window.

\subsection{Counter Attack}
`Counter Attack' is an in-possession transition phase. The ball is moved forward; possession changed within the previous $3.5$\,s (this timer is not increased while the phase is a counter attack); at least $5$ defenders recover; the ball speed exceeds $20$\,km/h; and no defender is closer to the ball than an attacker. The phase is kept alive until its criteria are not fulfilled for $400$\,ms within a $1.5$\,s window.

\subsection{Defensive Transition}
A transition follows a change of possession and is assigned to both teams: it is labeled an attacking transition for the team gaining possession and a defensive transition for the team losing it. The defensive transition requires that possession changed within the previous $5$\,s, that the ball is not yet under control, and that no other in-possession phase applies. The ball is considered under control if the distance to the closest attacker is less than $2.5$\,m, the distance to the closest defender is more than $4$\,m, and these conditions hold for $15$~consecutive frames. The phase is kept alive until its criteria are not fulfilled within $5$\,s; a defensive transition is additionally kept alive during a simultaneous counter-press until the end of that counter-press.

\section{Baseline GNN Layers}
\label{app:baseline-layers}
\paragraph{Graph Convolutions (GCNConv)}
The graph convolutional operator \citep{kipf_semi-supervised_2017} is based on the
symmetrically normalized adjacency matrix with self-loops. Let $\hat{A} = A + I$ denote
the adjacency matrix with added self-loops, and let $\hat{d}_i = \sum_j \hat{A}_{i,j}$
denote the degree of node $i$ including the self-loop. The GCN update rule can then be
written as
\begin{equation*}
    x'_i = \Theta^\top \sum_{j \in \mathcal{N}(i) \cup \{ i \}}
    \frac{\hat{A}_{j,i}}{\sqrt{\hat{d}_j \hat{d}_i}} \, x_j ,
\end{equation*}
where $\Theta$ is a trainable weight matrix.

GCNConv operates on a scalar-weighted adjacency matrix and, unlike SEAConv and GATConv, cannot incorporate the edge attribute vector $e_{i,j}$. In our experiments we therefore use the unweighted adjacency, i.e., $\hat{A}_{i,j} = 1$ for $(i,j) \in \mathcal{E}$ and $\hat{A}_{i,i} = 1$ for the added self-loops; the edge attributes are not used by this layer. The same holds for GINConv (see below), which aggregates neighbor features without edge information.
\paragraph{Graph Isomorphism Network (GINConv)}
Introduced by \citet{xu_how_2018}, message-passing within the graph isomorphism operator is written as 

\begin{equation*}
        x'_i = h_{\Theta}\left( (1 + \epsilon) \cdot x_i + \sum_{j \in \mathcal{N}(i)}x_j \right) ,
\end{equation*}
where $h_{\Theta}$ is a learnable multilayer perceptron (MLP) and $\epsilon$ a learnable parameter or a fixed scalar. 

\paragraph{Graph Attention (GATConv)}
\citet{velickovic_graph_2017} brought the concept of multi-head attention to GNNs \citep[see][for an introduction to the attention mechanism]{vaswani_attention_2017}.  We employ the GATv2 variant of \citet{brody_how_2022}, which reorders the attention computation such that the attention vector is applied \emph{after} the non-linearity, making the attention dynamic, i.e., conditional on the receiving node.
The updated node features $x'_i$ are calculated as
\begin{equation*}
    x'_i = \frac{1}{K}\sum_{k=1}^{K}
    \sum_{j \in \mathcal{N}(i) \cup \{i\}} \alpha^{(k)}_{i,j}
    \Theta_t^{(k)} x_j,
\end{equation*}
where $\Theta_t^{(k)}$ is a learnable weight matrix transforming the
message-sending node, unique per attention head $k$; the node's own
contribution enters the sum as a self-loop ($j = i$).
The attention coefficients $\alpha$ are computed as
\begin{equation*}
    \alpha_{i,j}^{(k)} =
    \frac{
        \exp\!\left(
            a^{(k)\top} \mathrm{LeakyReLU}\!\left(
                \Theta_s^{(k)} x_i +
                \Theta_t^{(k)} x_j +
                \Theta_e^{(k)} e_{i,j}
            \right)
        \right)
    }{
        \sum_{l \in \mathcal{N}(i) \cup \{i\}}
        \exp\!\left(
            a^{(k)\top} \mathrm{LeakyReLU}\!\left(
                \Theta_s^{(k)} x_i +
                \Theta_t^{(k)} x_l +
                \Theta_e^{(k)} e_{i,l}
            \right)
        \right)
    },
\end{equation*}
where $a^{(k)}$ is a learnable weight vector unique per head, $\Theta_s^{(k)}$ projects the receiving node, $e_{i,j}$ are edge attributes between nodes $i$ and $j$, and $\Theta_e^{(k)}$ is a learnable weight matrix that projects the edge attributes into the attention space. Note that the edge attributes modulate the attention weights but do not alter the messages themselves.

\section{Long Short-Term Memory Networks}
\label{sec:lstm}

A long short-term memory (LSTM) is a special type of Recurrent Neural Network (RNN), first introduced by \citet{hochreiter_long_1997}. RNNs are a subclass of neural networks designed to handle sequential data like text or time series. In contrast to feed-forward neural networks, they connect a neuron's output to the next input to account for the structural dependence of the data. As the unfolding of these networks leads to very deep structures, they typically suffer from the `vanishing gradient problem', which limits the learning ability of deep networks due to the vanishing of the weights' gradients through an increasing number of layers. LSTMs were designed to mitigate this issue, as gradients can flow through the network without being reduced too much. 

\begin{figure}[htbp]
    \centering
    \includegraphics[width=0.5\linewidth]{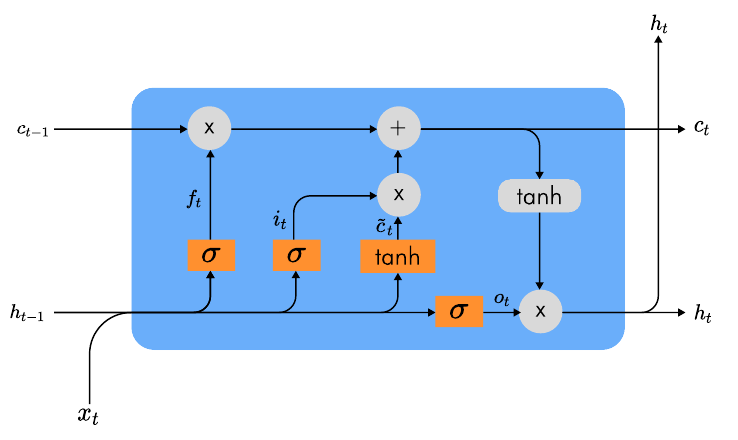}
    \caption{\textbf{LSTM cell}. Schematic visualization of an LSTM cell with input gate $i_t$, forget gate $f_t$, and output gate $o_t$ that use the previous cell state $c_{t-1}$ and output $h_{t-1}$ to update the cell state $c_t$ and output a new value $h_t$.}
    \label{fig:lstm-cell}
\end{figure}

For this purpose, LSTM cells possess a state as well as three gates --- an input, output, and forget gate. Figure \ref{fig:lstm-cell} visualizes such an LSTM cell. The cell state $c_t$ is updated in each cell based on the output of each of the three gates. As these gates are a single sigmoid layer, they essentially output how much information is added to the cell state, based on the previous state as well as this cell's input. The forget gate $f_t$ has the previous output $h_{t-1}$ and the current input $x_t$ as input and decides how much information is kept or otherwise forgotten. Additionally, the input gate $i_t$ determines how much of the input $x_t$ is used based on $h_{t-1}$, and multiplies this with the new candidate $\tilde{c}_t$ that is based on a $\tanh$ layer. The new cell state is then calculated as $c_t = f_t\cdot c_{t-1}+i_t\cdot\tilde{c}_t$. Finally, the output gate $o_t$ is responsible for how much of this cell state is kept as the cell's output $h_t$.

\section{Topological Features for the Baseline Model}
\label{app:topo-features}
In order to describe the spatial arrangement of players, topological features are derived from the player positions data for each second of observed match time. First, the convex hull of the observed team is computed as 
\begin{equation*}
    A_{\mathrm{team}} = \frac{1}{2} \left| \sum_{i \in \mathcal{V}_{\mathrm{team}}}(x_i y_{i+1} - x_{i+1}y_i) \right|,
\end{equation*}
where $\mathcal{V}_{\mathrm{team}}$ denotes the ordered vertices of the convex hull of the observed team's positions, with $A_{\mathrm{opp}}$ equivalently computed for the opponent. $x_i$ and $y_i$ are the x and y-coordinate of node $i$ \citep{preparata_computational_1985}.
The area ratio $A_{\mathrm{ratio}} = A_{\mathrm{team}} / A_{\mathrm{opp}}$ is used as a measure of relative spatial dominance. \citet{moura_spectral_2013} used similar features to describe team dynamics and tactics in football.

Additional topological features are computed via persistent homology. Concretely, we apply a Vietoris-Rips filtration \citep{edelsbrunner_computational_2010} to the combined 22-player spatial features $x^{\text{pos}} \in \mathbb{R}^{22 \times 2}$, obtaining persistence diagrams $\mathrm{dgm}_k$
for $k \in \{0, 1\}$. From these, the persistence entropy is derived \citep{chintakunta_entropy-based_2015} as 
\begin{equation*}
    E_{k}=-\sum_i\frac{\ell_i}{L} \log \left(\frac{\ell_i}{L}\right),
\end{equation*}
where $\ell_i=d_i-b_i$ is the lifetime of the
$i$-th feature and $L=\sum_i\ell_i$.
$E_0$ captures the entropy of connected-component merging (clustering structure), while $E_1$ captures the entropy of loops (encirclement/spatial holes between formations). The convex hulls are computed using the SciPy package \citep{virtanen_scipy_2020}, and the persistence diagrams and entropy are obtained using the Giotto-TDA package \citep{tauzin_giotto-tda_2020}.

\section{Cross-Entropy Loss}
\label{app:cross-entropy-loss}
The cross-entropy loss quantifies the divergence between a predicted probability distribution and the true class label, increasing as predictions diverge from the observed outcome. For binary classification with true label $y \in \{0, 1\}$ and predicted probability $\hat{y} \in [0, 1]$, the per-observation loss is defined as
\begin{equation*}
    L(y, \hat{y}) = -\bigl(y \cdot \log(\hat{y}) + (1 - y) \cdot \log(1 - \hat{y})\bigr).
\end{equation*}
For multiclass classification with $C > 2$ classes, let $y_c \in \{0, 1\}$ denote the one-hot indicator that the observation belongs to class $c$, and let $\hat{y}_c \in [0, 1]$ denote the corresponding predicted probability, with $\sum_{c=1}^{C} \hat{y}_c = 1$. The per-observation loss generalizes to
\begin{equation*}
    L(y, \hat{y}) = -\sum_{c=1}^{C} y_c \cdot \log(\hat{y}_c).
\end{equation*}
The negative sign ensures non-negativity, since $\log(\hat{y}_c) \leq 0$ for $\hat{y}_c \in (0, 1]$. 
The reported CE values in Section~\ref{sec:emp_res} correspond to the mean of $L$ over all observations in the validation set.

\section{Calibration Definitions}
\label{app:calibration-definitions}

A perfectly calibrated model predicts correct event probabilities. 
In the binary case, let $Y \in \{0,1\}$ denote the outcome and let $\hat{P} \in [0,1]$ be the model's predicted probability for $Y=1$; perfect calibration requires $\mathbb{P}(Y=1 \mid \hat{P}=p) = p$ for all $p \in [0,1]$.
As match phase prediction is a multiclass setting, this definition does not apply to the joint probability vector. 
Instead, marginal calibration is assessed by considering each class in a one-vs-rest view. 
Since $\mathbb{P}(Y=1 \mid \hat{P}=p)$ is unknown, we estimate it by binning class predictions into $M$ bins and, in each bin, comparing the observed fraction of positives to the mean predicted probability. This visualization is known as a reliability diagram.

The expected calibration error (ECE) is defined as
\begin{equation}
\label{eq:ece}
    \text{ECE} = \sum_{m=1}^{M}\frac{|B_m|}{n}|o_m - b_m|,
\end{equation}
with $B_m$ denoting bin $m$, $n$ the total number of observations in the validation set. Each sample enters every class's one-vs-rest problem. $o_m$ is the observed fraction of positives in bin $m$, and $b_m$ the mean predicted probability in bin $m$. The ECE is weighted by all observations, so for rare classes it is dominated by the many near-zero predictions and can't be compared across classes.

\section{Figures}
\begin{figure}[htbp]
    \centering
    \includegraphics[width=\linewidth]{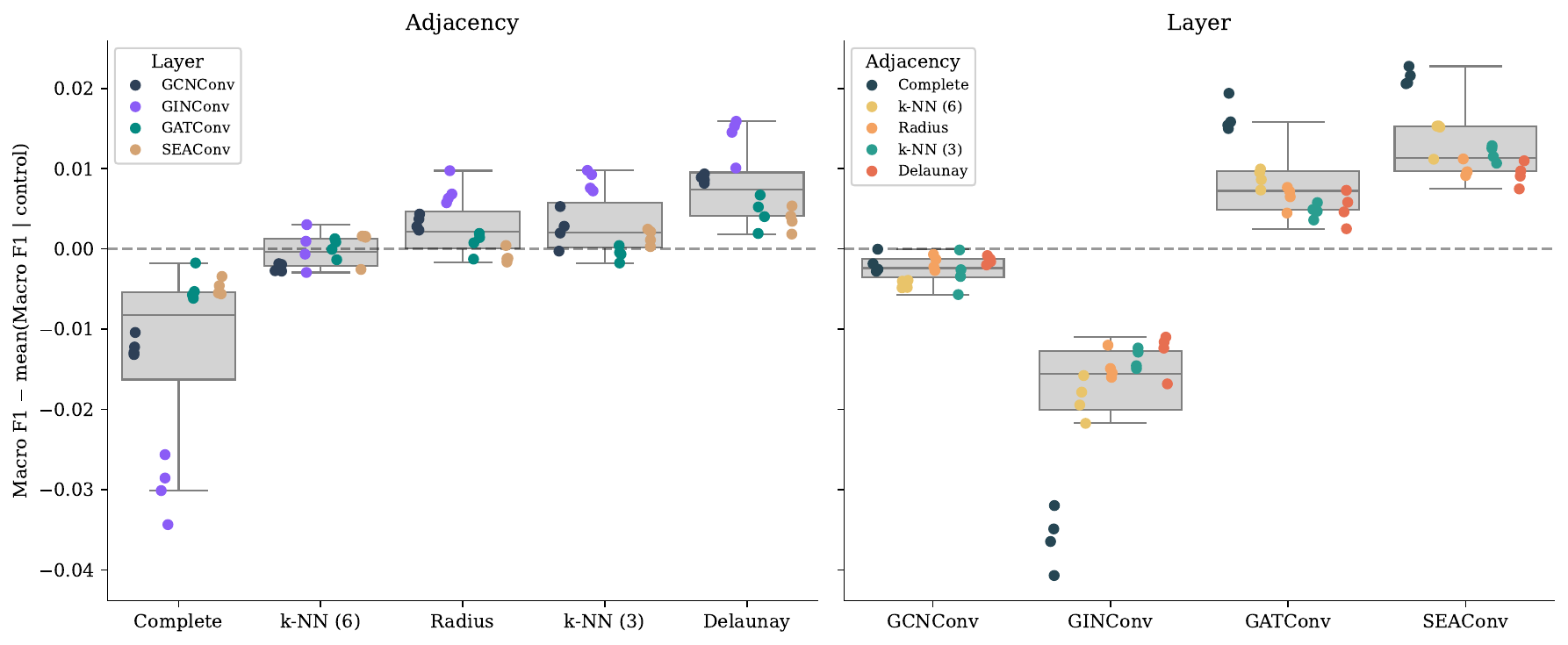}
    \caption{\textbf{Block-normalized residual validation macro F1}. Left: effect of adjacency after removing mean layer effects. Right: effect of layer after removing mean adjacency effects. Boxplots show residual distributions; colored points denote the control setting. Higher values indicate better performance.
}
    \label{fig:experiment_results_residuals}
\end{figure}

\clearpage
\section{Tables}
\begin{table}[htbp]
\caption{\textbf{Competitions and Seasons.} Number of matches in the data set by competition and season.}
\label{tab:competition_season_descriptive}
\centering
    \begin{tabular}{l @{\hskip 0.2in} cccccc @{\hskip 0.4in} r}
        \toprule
            &   \multicolumn{6}{c}{Season} & \\
        Competition & 19/20 & 20/21 & 21/22 & 22/23 & 23/24 & 24/25 & Total \\
        \midrule
        DFB-Pokal & 3 & 0 & 35 & 53 & 59 & 29 & 179 \\
        Senior national team & 0 & 0 & 0 & 3 & 11 & 4 & 18 \\
        Under-21 national team & 0 & 1 & 0 & 3 & 2 & 0 & 6 \\
        \midrule
        Total & 3 & 1 & 35 & 59 & 72 & 33 & 203 \\
        \bottomrule
    \end{tabular}
\end{table}

\begin{table}[htbp]
\centering
\caption{\textbf{Variables.} List of all variables contained in the data set with detailed explanations and processing methods.}
\label{tab:variables}
\begin{tabularx}{\textwidth}{l l X X}
\toprule
\textbf{Variable} & \textbf{Abbreviation} & \textbf{Explanation} & \textbf{Details} \\
\midrule
X-coordinate & X & Spatial horizontal coordinate of player / ball & Standardized to 0--1 based on pitch dimensions \\

Y-coordinate & Y & Spatial vertical coordinate of player / ball & Standardized to 0.176--0.824 to match the x-coordinate in aspect ratio for a \qtyproduct{105 x 68}{\metre} pitch\\

Speed & S & Movement speed of player / ball in m/s & Standardized based on a maximum value of \SI{10}{\metre\per\second} \\

Acceleration & A & Change in player speed in \SI{}{\metre\per\second\squared} & Standardized based on a maximum value of \SI{7}{\metre\per\second\squared} \\

Distance to Goal & DG & Distance to opponent goal & Computed from standardized coordinates \\

Distance to Ball & DB & Distance to the ball position & Computed from standardized coordinates \\

Team Association & TA & One-hot encoding of team association & Three columns indicating if the node is part of the observed team, opponent team, or the ball \\

Movement Direction & MD & X and Y deviation of player coordinates & Modeled using sine and cosine components \\

Angle to Goal & GA & Angle to opponent goal & Modeled using sine and cosine \\

Angle to Ball & BA & Angle to the ball position & Modeled using sine and cosine \\

\bottomrule
\end{tabularx}
\end{table}

\begin{table}[htbp]
\centering
\caption{\textbf{Hyperparameters.} Detailed collection of all hyperparameters used for model training. The column `Selection' contains the selection process and reasoning for the parameter value with references if the parameter is standard or common.}
\label{tab:hyperparameters}
\begin{tabularx}{\textwidth}{l l X}
        \toprule
        Hyperparameter & Value & Selection  \\
        \midrule
        Train/validation split & 70/30 (by matches) & - \\
        Loss function & Cross-entropy &  - \\
        Epochs & 100 (+ 10 GNN pre-training) & Computational constraints \\
        Truncated BPTT & 4 chunks per match & Computational constraints \\
        Gradient clipping & max-norm 1.0 & Common for RNNs \citep{pascanu_difficulty_2013} \\
        \midrule
        $r$ for radius graph & 0.2 ($\approx 20\,\mathrm{m}$ in physical units) & Manual \\
        \midrule
        Optimizer (graph encoder) & SGD (momentum 0.9) & Standard \citep{sutskever_importance_2013} \\
        Learning rate (graph encoder) & 0.01 & Common \\
        Optimizer (LSTM \& head) & AdamW & Standard \citep{kingma_adam_2015, loshchilov_decoupled_2019} \\
        Learning rate (LSTM \& head) & 0.001 & Standard \citep{kingma_adam_2015, loshchilov_sgdr_2017}  \\
        Weight decay (both) & $5 \times 10^{-4}$ & Standard in GNN literature \citep{kipf_semi-supervised_2017, velickovic_graph_2017} \\
        LR schedule (both) & Cosine annealing to $3 \times 10^{-6}$ & Standard technique \citep{loshchilov_sgdr_2017}, final value manually chosen \\
        \midrule
        Message-passing layers & 3 & Common, deeper GNNs oversmooth \citep{li_deeper_2018} \\
        Normalization (GNN) & PairNorm + SiLU after each layer & Design choice \\
        Dropout (GNN layers) & 0.1 & Standard \citep{vaswani_attention_2017} \\
        Dropout (LSTM) & 0.3 & Manual \\
        Attention heads & 4 & Manual\\
        Attention dropout & 0.1 & Standard \citep{vaswani_attention_2017}  \\
        Fourier $\sigma$ & $2\pi$ & Manual \\
        Hidden dimension (GNN) & 64 & Common \citep[e.g.][]{dwivedi_benchmarking_2023} \\
        LSTM layers & 1 & Manual \\
        Hidden dimension (LSTM) & 64 & Set in accordance with GNN \\
        \bottomrule
\end{tabularx}
\end{table}

\begin{table}[htbp]
    \centering
    \caption{\textbf{Pairwise Wilcoxon Signed-Rank Tests.} All pairwise comparisons of adjacency construction strategies (Panel~A) and message-passing layers (Panel~B) by Macro F1. Observations are paired by the complementary factor and seed ($n=16$ and $n=20$ matched pairs, respectively). $\Delta = \text{Macro F1(Method~A)} - \text{Macro F1(Method~B)}$; positive values favor Method~A. 95\% CI is a percentile bootstrap confidence interval on the mean paired difference. $p$-values are two-sided and Holm--Bonferroni corrected within each panel. \emph{Consistent} denotes the fraction of matched pairs where the sign of $\Delta$ of the individual pair is consistent with that of the mean.}
    \label{tab:wilcoxon_tests}
    \begin{tabular}{llcccc}
        \toprule
        & & \multicolumn{4}{c}{Macro F1} \\
        Method A & Method B & $\Delta$ & 95\% CI & $p$ (Holm) & Cons. \\
        \midrule
        \multicolumn{6}{l}{\textit{Panel A: Adjacency Construction} ($n=16$)} \\
        \midrule
        Delaunay & Complete & +0.0206 & [+0.0137,\;+0.0281] & 0.0003 & 16/16 \\
        k-NN ($k_{NN}=3$) & Complete & +0.0158 & [+0.0096,\;+0.0227] & 0.0003 & 16/16 \\
        Radius & Complete & +0.0154 & [+0.0091,\;+0.0222] & 0.0003 & 16/16 \\
        k-NN ($k_{NN}=6$) & Complete & +0.0125 & [+0.0077,\;+0.0180] & 0.0003 & 16/16 \\
        Delaunay & k-NN ($k_{NN}=6$) & +0.0081 & [+0.0058,\;+0.0104] & 0.0003 & 16/16 \\
        Delaunay & Radius & +0.0052 & [+0.0040,\;+0.0062] & 0.0003 & 15/16 \\
        Delaunay & k-NN ($k_{NN}=3$) & +0.0048 & [+0.0036,\;+0.0059] & 0.0003 & 16/16 \\
        k-NN ($k_{NN}=3$) & k-NN ($k_{NN}=6$) & +0.0033 & [+0.0015,\;+0.0052] & 0.0125 & 13/16 \\
        Radius & k-NN ($k_{NN}=6$) & +0.0029 & [+0.0009,\;+0.0049] & 0.0365 & 12/16 \\
        k-NN ($k_{NN}=3$) & Radius & +0.0004 & [-0.0007,\;+0.0015] & 0.4954 & 9/16 \\
        \midrule
        \multicolumn{6}{l}{\textit{Panel B: Message-Passing Layer} ($n=20$)} \\
        \midrule
        SEAConv & GINConv & +0.0325 & [+0.0271,\;+0.0386] & $1.1\times10^{-5}$ & 20/20 \\
        GATConv & GINConv & +0.0275 & [+0.0222,\;+0.0338] & $1.1\times10^{-5}$ & 20/20 \\
        GCNConv & GINConv & +0.0167 & [+0.0130,\;+0.0211] & $1.1\times10^{-5}$ & 20/20 \\
        SEAConv & GCNConv & +0.0158 & [+0.0137,\;+0.0180] & $1.1\times10^{-5}$ & 20/20 \\
        GATConv & GCNConv & +0.0108 & [+0.0088,\;+0.0129] & $1.1\times10^{-5}$ & 20/20 \\
        SEAConv & GATConv & +0.0050 & [+0.0042,\;+0.0060] & $1.1\times10^{-5}$ & 20/20 \\
        \bottomrule
    \end{tabular}
\end{table}

\begin{table}[htbp]
    \centering
    \caption{\textbf{Class-wise Evaluation Metrics.} Per-class metrics of the GNN's predictions on the validation set, averaged over the four seeds of the model using Delaunay adjacency and SEAConv layers, each evaluated at its final checkpoint (epoch 100). CE is the cross-entropy conditional on the true class, $\text{CE}_c = -\frac{1}{n_c}\sum_{i:\,y_i=c}\log \hat{y}_{i,c}$. ECE (Equation~\ref{eq:ece}, 10 equal-width bins) and F1 are computed one-vs-rest, taking the class with the highest predicted probability as the prediction for F1.}
    \label{tab:calibration_metrics}
    \begin{tabular}{ll@{\hskip 0.4in}ccc}
        \toprule
        Class & Class Size &  CE Loss & ECE & F1 \\
        \midrule
        No Phase & 56.3\% & 0.153 & 0.0061 & 0.936 \\
        Build Up & 16.2\% & 0.204 & 0.0051 & 0.912 \\
        Low Block & 2.1\% & 0.422 & 0.0018 & 0.811 \\
        Mid Block & 16.4\% & 0.147 & 0.0015 & 0.932 \\
        High Block & 1.0\% & 0.591 & 0.0006 & 0.795 \\
        Defensive Transition & 5.2\% & 0.846 & 0.0012 & 0.659 \\
        Counter Attack & 2.9\% & 1.371 & 0.0041 & 0.535 \\
        \bottomrule
    \end{tabular}
\end{table}

\begin{table}[htbp]
\centering
\caption{\textbf{Global State and Match Phase.} Empirical conditional distributions of global state $u_t$ (ball in play, own team in possession) and the annotated phase, over the 689{,}570 validation observations used in the attribution analysis. Phases are No Phase, Build Up (BU), Low Block (LB), Mid Block (MB), High Block (HB), Defensive Transition (DT) and Counter Attack (CA).}
\label{tab:global-state-distribution}
\smallskip
\textit{Panel A: phase distribution given the global state}
\smallskip
\begin{tabular}{lrrrrrrrr}
\toprule
Global state & No Phase & BU & LB & MB & HB & DT & CA & $n$ \\
\midrule
In play, own poss. & 36.1\% & 52.8\% & 0.0\% & 0.0\% & 0.0\% & 2.3\% & 8.8\% & 204{,}378 \\
In play, opp. poss. & 21.9\% & 0.3\% & 6.6\% & 53.1\% & 3.2\% & 14.2\% & 0.6\% & 206{,}939 \\
Dead ball, own poss. & 99.1\% & 0.7\% & 0.0\% & 0.0\% & 0.0\% & 0.0\% & 0.2\% & 133{,}602 \\
Dead ball, opp. poss. & 94.4\% & 1.4\% & 0.5\% & 2.1\% & 0.1\% & 1.1\% & 0.5\% & 144{,}651 \\
\bottomrule
\end{tabular}

\bigskip
\textit{Panel B: global-state distribution given the phase}
\smallskip
\begin{tabular}{lrrrrr}
\toprule
 & \multicolumn{2}{c}{Ball in play} & \multicolumn{2}{c}{Dead ball} & \\
\cmidrule(lr){2-3}\cmidrule(lr){4-5}
Phase & Own poss. & Opp.\ poss. & Own poss. & Opp.\ poss. & $n$ \\
\midrule
No Phase & 19.0\% & 11.7\% & 34.1\% & 35.2\% & 388{,}192 \\
Build Up & 96.9\% & 0.5\% & 0.8\% & 1.8\% & 111{,}453 \\
Low Block & 0.0\% & 95.3\% & 0.0\% & 4.7\% & 14{,}391 \\
Mid Block & 0.0\% & 97.4\% & 0.0\% & 2.6\% & 112{,}936 \\
High Block & 0.0\% & 98.1\% & 0.0\% & 1.9\% & 6{,}745 \\
Def. Transition & 13.2\% & 82.2\% & 0.1\% & 4.5\% & 35{,}783 \\
Counter Attack & 89.2\% & 6.2\% & 1.1\% & 3.5\% & 20{,}070 \\
\bottomrule
\end{tabular}
\end{table}

\clearpage
\begin{longtable}{lllrrrrr}
    \caption{\textbf{Full Results.} Detailed results and validation metrics for all combinations run. The table is sorted by Macro F1.}
    \label{tab:full_results} \\
    \toprule    
    Adjacency Construction & Layer & Seed & Accuracy & Bal. Accuracy & CE Loss & Macro F1 & Weighted F1 \\
    \midrule
    \endfirsthead

    \multicolumn{8}{l}{\textit{Continued from previous page}} \\
    \toprule
    Adjacency Construction & Layer & Seed & Accuracy & Bal. Accuracy & CE Loss & Macro F1 & Weighted F1 \\
    \midrule
    \endhead

    \midrule
    \multicolumn{8}{r}{\textit{Continued on next page}} \\
    \endfoot

    \bottomrule
    \endlastfoot
Delaunay & SEAConv & 123 & 0.9039 & 0.7871 & 0.2397 & 0.7988 & 0.9017 \\
Delaunay & SEAConv & 789 & 0.9034 & 0.7872 & 0.2408 & 0.7976 & 0.9010 \\
Delaunay & SEAConv & 456 & 0.9038 & 0.7837 & 0.2406 & 0.7969 & 0.9011 \\
k-NN ($k_{NN}=3$) & SEAConv & 123 & 0.9034 & 0.7838 & 0.2422 & 0.7959 & 0.9006 \\
k-NN ($k_{NN}=3$) & SEAConv & 456 & 0.9034 & 0.7818 & 0.2410 & 0.7956 & 0.9007 \\
Delaunay & SEAConv & 42 & 0.9032 & 0.7830 & 0.2430 & 0.7953 & 0.9008 \\
Delaunay & GATConv & 789 & 0.9036 & 0.7870 & 0.2401 & 0.7951 & 0.9016 \\
k-NN ($k_{NN}=6$) & SEAConv & 456 & 0.9029 & 0.7828 & 0.2445 & 0.7951 & 0.9003 \\
k-NN ($k_{NN}=6$) & SEAConv & 123 & 0.9028 & 0.7802 & 0.2437 & 0.7950 & 0.9002 \\
k-NN ($k_{NN}=6$) & SEAConv & 789 & 0.9031 & 0.7818 & 0.2435 & 0.7949 & 0.9004 \\
k-NN ($k_{NN}=3$) & SEAConv & 789 & 0.9029 & 0.7808 & 0.2413 & 0.7946 & 0.9002 \\
Radius & SEAConv & 42 & 0.9016 & 0.7813 & 0.2477 & 0.7939 & 0.8991 \\
k-NN ($k_{NN}=3$) & SEAConv & 42 & 0.9021 & 0.7805 & 0.2438 & 0.7938 & 0.8995 \\
Delaunay & GATConv & 456 & 0.9037 & 0.7802 & 0.2406 & 0.7937 & 0.9013 \\
Delaunay & GATConv & 123 & 0.9030 & 0.7826 & 0.2418 & 0.7925 & 0.9007 \\
Radius & SEAConv & 456 & 0.9016 & 0.7788 & 0.2479 & 0.7923 & 0.8988 \\
Radius & SEAConv & 789 & 0.9005 & 0.7784 & 0.2482 & 0.7922 & 0.8978 \\
Radius & SEAConv & 123 & 0.9011 & 0.7778 & 0.2499 & 0.7918 & 0.8983 \\
k-NN ($k_{NN}=6$) & SEAConv & 42 & 0.9013 & 0.7758 & 0.2463 & 0.7909 & 0.8985 \\
Radius & GATConv & 42 & 0.9023 & 0.7795 & 0.2465 & 0.7904 & 0.9000 \\
Delaunay & GATConv & 42 & 0.9030 & 0.7764 & 0.2416 & 0.7904 & 0.9005 \\
Complete & SEAConv & 123 & 0.8997 & 0.7794 & 0.2616 & 0.7900 & 0.8972 \\
Radius & GATConv & 123 & 0.9021 & 0.7801 & 0.2486 & 0.7899 & 0.8998 \\
k-NN ($k_{NN}=6$) & GATConv & 456 & 0.9019 & 0.7738 & 0.2475 & 0.7897 & 0.8993 \\
k-NN ($k_{NN}=6$) & GATConv & 123 & 0.9017 & 0.7766 & 0.2473 & 0.7893 & 0.8995 \\
Radius & GATConv & 789 & 0.9022 & 0.7743 & 0.2461 & 0.7892 & 0.8998 \\
Complete & SEAConv & 456 & 0.8998 & 0.7780 & 0.2605 & 0.7889 & 0.8970 \\
k-NN ($k_{NN}=3$) & GATConv & 42 & 0.9017 & 0.7765 & 0.2458 & 0.7888 & 0.8992 \\
k-NN ($k_{NN}=6$) & GATConv & 42 & 0.9021 & 0.7723 & 0.2470 & 0.7884 & 0.8993 \\
Complete & SEAConv & 42 & 0.8991 & 0.7788 & 0.2628 & 0.7880 & 0.8965 \\
k-NN ($k_{NN}=3$) & GATConv & 456 & 0.9018 & 0.7721 & 0.2463 & 0.7880 & 0.8992 \\
Complete & SEAConv & 789 & 0.8997 & 0.7763 & 0.2614 & 0.7879 & 0.8969 \\
k-NN ($k_{NN}=3$) & GATConv & 789 & 0.9019 & 0.7740 & 0.2439 & 0.7878 & 0.8991 \\
Radius & GATConv & 456 & 0.9017 & 0.7724 & 0.2476 & 0.7872 & 0.8991 \\
k-NN ($k_{NN}=6$) & GATConv & 789 & 0.9019 & 0.7703 & 0.2474 & 0.7871 & 0.8992 \\
Delaunay & GCNConv & 789 & 0.9005 & 0.7735 & 0.2460 & 0.7870 & 0.8974 \\
k-NN ($k_{NN}=3$) & GATConv & 123 & 0.9008 & 0.7729 & 0.2487 & 0.7867 & 0.8983 \\
Complete & GATConv & 456 & 0.9011 & 0.7716 & 0.2523 & 0.7867 & 0.8985 \\
Delaunay & GCNConv & 42 & 0.8999 & 0.7744 & 0.2476 & 0.7866 & 0.8969 \\
Delaunay & GCNConv & 123 & 0.8997 & 0.7724 & 0.2487 & 0.7863 & 0.8969 \\
Delaunay & GCNConv & 456 & 0.9003 & 0.7698 & 0.2480 & 0.7858 & 0.8971 \\
Complete & GATConv & 789 & 0.8998 & 0.7683 & 0.2558 & 0.7831 & 0.8974 \\
k-NN ($k_{NN}=3$) & GCNConv & 789 & 0.8989 & 0.7712 & 0.2518 & 0.7829 & 0.8961 \\
Complete & GATConv & 123 & 0.9001 & 0.7650 & 0.2555 & 0.7827 & 0.8974 \\
Complete & GATConv & 42 & 0.8997 & 0.7687 & 0.2577 & 0.7823 & 0.8972 \\
Radius & GCNConv & 42 & 0.8978 & 0.7685 & 0.2560 & 0.7820 & 0.8951 \\
Radius & GCNConv & 456 & 0.8983 & 0.7661 & 0.2546 & 0.7814 & 0.8952 \\
k-NN ($k_{NN}=3$) & GCNConv & 42 & 0.8979 & 0.7680 & 0.2539 & 0.7805 & 0.8949 \\
Radius & GCNConv & 789 & 0.8985 & 0.7636 & 0.2549 & 0.7804 & 0.8952 \\
Radius & GCNConv & 123 & 0.8982 & 0.7608 & 0.2546 & 0.7800 & 0.8948 \\
k-NN ($k_{NN}=3$) & GCNConv & 123 & 0.8975 & 0.7656 & 0.2558 & 0.7796 & 0.8944 \\
k-NN ($k_{NN}=3$) & GCNConv & 456 & 0.8976 & 0.7580 & 0.2537 & 0.7774 & 0.8941 \\
Delaunay & GINConv & 123 & 0.8979 & 0.7575 & 0.2609 & 0.7768 & 0.8951 \\
Delaunay & GINConv & 789 & 0.8976 & 0.7526 & 0.2642 & 0.7762 & 0.8946 \\
k-NN ($k_{NN}=6$) & GCNConv & 42 & 0.8964 & 0.7581 & 0.2584 & 0.7758 & 0.8932 \\
k-NN ($k_{NN}=6$) & GCNConv & 789 & 0.8970 & 0.7577 & 0.2578 & 0.7757 & 0.8936 \\
Delaunay & GINConv & 42 & 0.8974 & 0.7558 & 0.2659 & 0.7755 & 0.8946 \\
k-NN ($k_{NN}=6$) & GCNConv & 456 & 0.8961 & 0.7557 & 0.2599 & 0.7749 & 0.8930 \\
k-NN ($k_{NN}=6$) & GCNConv & 123 & 0.8962 & 0.7576 & 0.2609 & 0.7749 & 0.8929 \\
Delaunay & GINConv & 456 & 0.8969 & 0.7483 & 0.2679 & 0.7710 & 0.8939 \\
k-NN ($k_{NN}=3$) & GINConv & 789 & 0.8949 & 0.7499 & 0.2706 & 0.7707 & 0.8920 \\
Radius & GINConv & 789 & 0.8941 & 0.7538 & 0.2787 & 0.7707 & 0.8912 \\
k-NN ($k_{NN}=3$) & GINConv & 123 & 0.8955 & 0.7490 & 0.2699 & 0.7702 & 0.8927 \\
k-NN ($k_{NN}=3$) & GINConv & 456 & 0.8954 & 0.7440 & 0.2753 & 0.7685 & 0.8922 \\
k-NN ($k_{NN}=3$) & GINConv & 42 & 0.8951 & 0.7398 & 0.2722 & 0.7682 & 0.8916 \\
Radius & GINConv & 42 & 0.8938 & 0.7431 & 0.2830 & 0.7678 & 0.8903 \\
Complete & GCNConv & 123 & 0.8927 & 0.7495 & 0.2726 & 0.7672 & 0.8896 \\
Radius & GINConv & 123 & 0.8941 & 0.7452 & 0.2802 & 0.7672 & 0.8907 \\
Radius & GINConv & 456 & 0.8934 & 0.7457 & 0.2849 & 0.7667 & 0.8899 \\
Complete & GCNConv & 42 & 0.8922 & 0.7495 & 0.2719 & 0.7654 & 0.8889 \\
Complete & GCNConv & 456 & 0.8924 & 0.7461 & 0.2720 & 0.7647 & 0.8892 \\
Complete & GCNConv & 789 & 0.8924 & 0.7481 & 0.2728 & 0.7645 & 0.8889 \\
k-NN ($k_{NN}=6$) & GINConv & 123 & 0.8932 & 0.7391 & 0.2781 & 0.7640 & 0.8902 \\
k-NN ($k_{NN}=6$) & GINConv & 42 & 0.8938 & 0.7355 & 0.2776 & 0.7619 & 0.8904 \\
k-NN ($k_{NN}=6$) & GINConv & 789 & 0.8922 & 0.7327 & 0.2809 & 0.7603 & 0.8888 \\
k-NN ($k_{NN}=6$) & GINConv & 456 & 0.8924 & 0.7313 & 0.2829 & 0.7580 & 0.8892 \\
Complete & GINConv & 456 & 0.8831 & 0.7089 & 0.3146 & 0.7353 & 0.8797 \\
Complete & GINConv & 42 & 0.8837 & 0.7014 & 0.3146 & 0.7324 & 0.8801 \\
Complete & GINConv & 789 & 0.8827 & 0.7032 & 0.3192 & 0.7308 & 0.8789 \\
Complete & GINConv & 123 & 0.8827 & 0.6956 & 0.3187 & 0.7266 & 0.8788 \\
\end{longtable}

\end{document}